\magnification=\magstep1 
\overfullrule=0pt
\font\huge=cmr10 scaled \magstep2
 
\font\smal=cmr7       

\def\i{{\rm i}\,}

\catcode`\@=11
\font\tenmsa=msam10
\font\sevenmsa=msam7
\font\fivemsa=msam5
\font\tenmsb=msbm10
\font\sevenmsb=msbm7
\font\fivemsb=msbm5
\newfam\msafam
\newfam\msbfam
\textfont\msafam=\tenmsa  \scriptfont\msafam=\sevenmsa
  \scriptscriptfont\msafam=\fivemsa
\textfont\msbfam=\tenmsb  \scriptfont\msbfam=\sevenmsb
  \scriptscriptfont\msbfam=\fivemsb
\def\hexnumber@#1{\ifcase#1 0\or1\or2\or3\or4\or5\or6\or7\or8\or9\or
	A\or B\or C\or D\or E\or F\fi }

\def\Bbb{\ifmmode\let\next\Bbb@\else
 \def\next{\errmessage{Use \string\Bbb\space only in math mode}}\fi\next}
\def\Bbb@#1{{\Bbb@@{#1}}}
\def\Bbb@@#1{\fam\msbfam#1}
\def\Z{{\Bbb Z}} \def\Q{{\Bbb Q}}  \def\C{{\Bbb C}}  \def\R{{\Bbb R}}
\def\L{\Lambda}  \def\G{\Gamma}  \def\la{\lambda}  \def\z{{\cal Z}}
\def\I{{\cal I}}  \def\p{{\cal P}} \def\k{{\overline{k}}}
\def\D{{\cal D}}  \def\E{{\cal E}}  \def\pP{{\overline{P}}}
\def\ch{{\overline{\chi}}}

{\nopagenumbers
\bigskip\bigskip
\centerline{{\bf \huge  U(1)$^{ m}$ Modular Invariants, N=2 Minimal Models,}}
\medskip\centerline{{\bf\huge and the Quantum Hall Effect}}
\bigskip\bigskip\centerline{{\bf Terry Gannon\footnote{$^{\dag}$}{{\smal 
Address from September 1996:}} \footnote{}{{\smal 
Dept of Math, York University, 4700 Keele Street, North York, Canada M3J 
1P3}}  \footnote{}{{\smal e-mail: tgannon@mathstat.yorku.ca}}}} \medskip
\centerline{{\it Max-Planck-Institut f\"ur Mathematik,}}
\centerline{{\it  Gottfried-Claren-Strasse 26,}} 
\centerline{{\it 53225 Bonn, Germany}}
\bigskip \bigskip\bigskip
\centerline{{\bf Abstract}}\bigskip
The problem of finding all possible effective field theories for the
quantum Hall effect is closely related to the problem of classifying all
possible modular invariant partition functions for the algebra $\widehat{
u(1)^{\oplus m}}$, as was argued recently by Cappelli and Zemba. This latter
problem is also a natural one from the perspective of conformal field
theory. In this paper
we completely solve this problem, expressing the answer in terms of self-dual
lattices, or equivalently, rational points on the dual Grassmannian
$G_{m,m}(\R)^*$. We also find all modular invariant partition functions for 
affine $su(2)\oplus u(1)^{\oplus m}$, from which we obtain the classification
of all $N=2$ superconformal minimal models. The `A-D-E classification'
of these, though often quoted in the literature, turns out to be a very 
coarse-grained one: e.g.\ associated with the names $E_6,E_7,E_8$, 
respectively, are precisely 20,30,24 different partition functions. As a
by-product of our analysis, we find that the list
of modular invariants for $\widehat{su(2)}$ lengthens surprisingly little 
when commutation with $T$ -- i.e.\ invariance under 
$\tau\mapsto\tau+1$ -- is ignored: the other conditions are far more
essential.\vfill \eject}  \pageno=1

\centerline{{\bf 1. Introduction}}\bigskip 

The quantum Hall effect for 2-dimensional incompressible quantum fluids
has received considerable attention in recent years, both from theorists
and experimentalists (see e.g.\ [1]). First observed experimentally in 
the early 1980s, a major theoretical step was done by Laughlin and his
theory of plateaux. Experimentally, one observes universality -- i.e.\
some features of the effect, e.g.\ the possible values of the filling
factor $\nu$, are largely independent of impurities and geometry, for
example.

One is thus led to study universality classes of incompressible quantum
Hall fluids by effective field theories, in the long-distance/low-temperature
limit. There are at least two main approaches to this. One (see e.g.\
[2] and references therein) starts with an abelian Chern-Simons theory,
while the other (see e.g.\ [3] and references therein) expresses
incompressibility algebraically and investigates $W_{1+\infty}$ conformal
field theories. The two approaches are related, and in recent work [4]
(see also [5]) proposed looking at modular invariant partition functions for 
these theories.
Both these approaches correspond to looking at modular invariant
sesquilinear combinations of the characters of the affine algebra $U_m:=
\widehat{u(1)^{\oplus m}}$ at some (matrix-valued) level $k$. 

There is a family of rational conformal field theories (RCFTs) for each 
choice of current(=nontwisted affine 
Kac-Moody) algebra $g$ (see e.g.\ [6] and references therein for a review of 
this problem), and the
choice $g=U_m$ is a natural one from this perspective as well. In this paper 
we find all such partition functions. The solution has a simple geometric
description in terms of self-dual lattices, or equivalently rational
points on the dual of the Grassmannian $G_{m,m}(\R)$. The theories in [2] 
correspond to a 
small subset of these, namely the diagonal partition functions. [4] have
suggested that some of the non-diagonal partition functions provide a natural
explanation  of some of the plateaux falling out of the Jain sequence, which
have been experimentally observed (e.g.\ $\nu=4/11$).

Of course there are long-established connections between lattices and the 
quantum Hall effect (see e.g.\ [2]). The difference here is that the lattices
are all self-dual, and have dimension $2m$ (instead of $m$).

As is well-known, the quantum Hall theorists are plagued by the difficulty 
of having too many possible effective field theories to choose from -- far 
more than have been observed
experimentally. What still seems to be missing is an understanding of the
stability, i.e.\ width, of the plateaux -- it appears that 
only heuristic proposals  in this direction
have so far been made. This short paper cannot contribute to this difficult
problem, except indirectly by providing a complete list of the possible 
effective theories (more precisely, a complete list of the possible partition
functions, which determine all possible spectra of these theories).

The second classification we obtain in this paper is that of the $N=2$ 
superconformal
minimal models. The conformal (i.e.\ $N=0$) minimal models are classified
in [7], and the $N=1$ ones in [8]. The $N=2$ super-Virasoro algebra
is of great interest because $N=1$ space-time supersymmetry in string theory
is related to $N=2$ world-sheet supersymmetry (e.g.\ Gepner [9] uses the $N=2$
minimal models to compactify the heterotic string), and also because
of the possible relation of the $N=2$ models with Calabi-Yau manifolds and 
with Landau-Ginzburg theories. The classification of the $N=2$ minimal
models has been addressed many times in the literature (see e.g.\ [10,9]),
and it is often claimed that they fall into the A-D-E pattern well known 
to conformal field theorists. Here we give the  first rigourous and
complete classification of the possible $N=2$ minimal model partition 
functions. Previous attempts generally assume some sort of 
factorisation at the level of the individual partition functions
themselves,
an assumption which is simply wrong. Thus we find many more
partition functions, and unfortunately there seems to be
no natural relation between our list and the A-D-E pattern.

The relation between the effective field theories for
the quantum Hall effect, and the $N=2$ minimal models,
is that their classifications reduce to the modular invariant classifications 
of  $U_m$ and $\hat{A_1}\oplus U_2$, respectively, and the techniques used
to solve $U_m$ help to solve $\hat{A_1}\oplus U_m$.

The activity at present concerning the classification of modular invariants is 
following a clearly defined program (see e.g.\ [6]) aiming at achieving this 
classification for all {\it simple} affine algebras. The present paper falls
outside this program. Its justification is that it accomplishes the
classification for two infinite families of (non-simple) algebras, both 
of which concern problems of immediate physical interest.

\bigskip\bigskip\centerline{{\bf 2. Background}}\bigskip

The notions of {\it lattice} $\L$, its {\it dual} $\L^*$, and its {\it 
determinant} $|\L|$, 
are well-known. An {\it integral} lattice obeys $\L\subset \L^*$,
and a {\it self-dual} one obeys $\L=\L^*$. Equivalently, $\L$ is self-dual iff
it is integral and has determinant $|\L|=1$. An integral lattice is {\it even} 
if all its norms $x^2$ are even, otherwise it is called {\it odd}. The 
operation $\oplus$ denotes orthogonal direct sum. See e.g.\ Chapter 2 of
[11] for definitions.

An RCFT possesses a finite set $P_+$ of labels (weights), and a 
complex-valued function (character) $\chi_a$ for each $a\in P_+$. The 
modular group $SL_2(\Z)$ acts on these $\chi_a$:
$$\eqalignno{\left(\matrix{0&-1\cr 1&0}\right)\cdot \chi_a=&\sum_{b\in P_+}
S_{a,b}\,\chi_b&(2.1a)\cr
\left(\matrix{1&1\cr 0&1}\right)\cdot\chi_a=&\sum_{b\in P_+}
T_{a,b}\,\chi_b\ .&(2.1b)\cr}$$
$S$ and $T$ are unitary and symmetric, and $T$ is diagonal. There is a 
distinguished weight $0\in P_+$ for which 
$$S_{0,a}\ge S_{0,0}>0\ .\eqno(2.1c)$$
The spectrum of the RCFT is encoded in its (genus 1) partition function
$${{\cal Z}}=\sum_{a,b\in P_+}M_{a,b}\,{\chi}_a\,{\chi}^*_b\ .\eqno(2.2a)$$
(Strictly speaking, the $a$ and $b$ in (2.2a) may come from different sets
$P_+^L$, $P_+^R$, respectively -- such $M$ are called {\it heterotic} and do 
occur in this paper. For notational simplicity we will usually ignore this
technicality. For example, we will never write e.g.\ $S^L$ or $0_L$ -- no 
confusion should ever result.) The coefficient matrix $M$ obeys
$$\eqalignno{M_{a,b}\in&\,\Z_{\ge}\qquad \forall a,b\in P_+\ ,&(2.2b)\cr
M_{0,0}=&\,1\ .&(2.2c)\cr}$$
Usually in a RCFT one requires invariance of $\z$ under the full modular group
$SL_2(\Z)$: 
$$\eqalignno{SM=&\,MS&(2.3a)\cr TM=&\,MT\ .&(2.3b)\cr}$$ 
By {\it physical invariant}  is meant any matrix $M$, or equivalently
the corresponding function $\z$ in (2.2a), obeying (2.2b), (2.2c), (2.3a) and 
(2.3b). We will use the term {\it weak invariant} to denote
any  $M$ (or  $\z$) obeying (2.2b), (2.2c) and (2.3a). In this paper we
classify all the physical/weak invariants for certain choices of $\chi_a$,
motivated by the quantum Hall effect and the $N=2$ super-Virasoro algebra.
The physical invariant classification for other $\chi_a$ has been the
subject of much work -- see e.g.\ [7,8,6] and references therein.

Incidently, equations (2.3) require the $\chi_a$ to be linearly independent.
This is usually accomplished in practice by giving them full variable
dependence (i.e.\ including zero-mode oscillations).

A rich source of RCFT data ($P_+,\,\{\chi_a\})$ comes from the representations 
of affine algebras.
The representation theory of the affine algebra $U_m=\widehat{u(1)^{\oplus 
m}}$ at level $k$ (more concisely, $U_{m,k}$) is well-known -- see e.g.\ 
Ch.\ 12 of [12]. $k$ here is an
$m\times m$ symmetric integral matrix -- it is common to call it `level' 
by analogy with the other affine algebras although its nature here is a
little different. (Our convention for $U_m$ level is double what other 
writers often choose.) 
Let $\G_k$ be the corresponding 
integral lattice, i.e.\ it will have a basis $\{e_1,\ldots,e_m\}$ satisfying
$e_i\cdot e_j=k_{ij}$. Let $\{e_1^*,\ldots,e_m^*\}$ be the corresponding
dual basis: $e_i\cdot e_j^*=\delta_{i,j}$. There is an integrable 
representation of $U_m$ for each choice
of highest weight $\la\in \pP^k:=\G_k^*/\G_k$; its character $\ch_\la^k(\tau,z)$
is proportional to the $m$-dimensional theta function $\theta_{\la}(z|\tau)$,
where the proportionality constant is independent of $\la$, and where
$$\theta_S(z|\tau)=\sum_{x\in S}\exp[\pi \i \tau x^2+2\pi \i z\cdot x]$$
for any set $S\subset \Q\otimes \G_k$. Here, $z\in \C\otimes \G_k$, $\tau
\in\C\,$; when $S$ is the translate of an $m$-dimensional lattice, $\theta_S$ 
will converge for Im$(\tau)>0$
provided the lattice is Euclidean, i.e.\ provided
in our case $k$ is positive definite.

The simplest case is when $k$ is diagonal, in which case $\Gamma_k=(\sqrt{
k_{11}}\,\Z)\oplus\cdots\oplus(\sqrt{k_{mm}}\,\Z)$, $\G_k^*=(\sqrt{{1\over
k_{11}}}\Z)\oplus\cdots\oplus(\sqrt{{1\over k_{mm}}}\Z)$, $e_i^*=e_i/k_{ii}$, 
and $\theta_{\la}$ reduces to a product of 1-dimensional theta functions.
We will usually denote a weight $\la\in\pP^k$ for such $k$ by its (integer)
components with respect to this dual basis $e_i^*$.

Strictly speaking, highest weight representations of $U_m$ require 
$k$ to be diagonal (as well as positive definite and integral). However all of
our formulas and arguments are independent of this restriction. Moreover,
the more important structure for RCFT is the chiral algebra(=vertex operator
algebra), which is well-defined for any positive definite, integral $k$
(non-integral $k$ would correspond to irrational CFTs).

We will thus assume throughout this paper that the level $k$ is positive 
definite and integral, though not
necessarily diagonal. Positive-definiteness is necessary for convergence of 
the partition
function, and hence for the existence of a RCFT. Physically (i.e.\ for
the quantum Hall effect discussed below), this would
correspond to all the excitations on each edge having equal chirality;
the more general situation (of mixed chiralities) can be easily accommodated
within this picture by using the following recipe: first find an orthogonal
sublattice of $\G_k$, using Gram-Schmidt -- i.e.\ find independent vectors
$v_i\in\Z e_1+\cdots+\Z e_i$, $i=1,\ldots,m$, such that $v_i\cdot v_j=0$
for $i\ne j$. Provided each $v_i^2\ne 0$, define $k'$ to be the diagonal
matrix $(|v_1^2|)\oplus\cdots\oplus(|v_m^2|)$. What we have effectively
done is moved all the excitations with wrong chirality to the opposite
edge. The original mixed chirality theory will then be constructable from
one of those at level $k'$ by returning all the excitations to their
proper edge. This is precisely what we do below with e.g.\ $U_1\oplus \hat{
A}_{m-1,1}$ theories -- see (2.6)  -- as well as in the correspondence 
between $N=2$
minimal models and $\hat{A}_1\oplus U_2$ theories in (6.1d). This recipe 
breaks down when some of the $v_i^2=0$,
but for such a case it would be doubtful that the theory could correspond 
naturally to a RCFT.

The modular transformation properties of $\overline{\chi}^k_\la$ are given by 
the matrices
$$\eqalignno{\overline{T}_{\la,\mu}^2=&\delta_{\la,\mu}\exp[2\pi \i(\la^2-
{m\over 12})]\ ,&(2.4a)\cr
\overline{S}_{\la,\mu}=&{1\over \sqrt{|k|}}\exp[-2\pi \i\la\cdot \mu]\ .&
(2.4b)\cr}$$
These correspond to the transformations $(\tau,z)\mapsto(\tau+2,z)$, and
$(\tau,z)\mapsto (-1/\tau, z/\tau)$, respectively (in this second 
transformation
we are ignoring a multiplicative factor which is not important for our purposes). 
Both matrices are unitary and
symmetric. We use the notation $\overline{T}^2$ purely formally here -- its 
square-root $\overline{T}$ will exist iff each $k_{ii}$ is even.

The partition function $\overline{{\cal Z}}$ built from these $\overline{
\chi}_\la^k$ enters naturally into the 
classification problem of effective field theories for incompressible 
quantum Hall fluids for generic hierarchical plateaux, where it describes
the pairings of excitations on the two edges of an annulus -- see [4,5]
for a discussion. For example, $m={\rm dim}\,\G_k$ corresponds to the
number of independent bosons, i.e.\ edge currents, and equals the central
charge of the RCFT ($m=1$ for Laughlin fluids). There are two main differences
introduced here from the generic RCFT situation. One is that for quantum Hall 
fluids equation (2.3b) should be weakened to
$$ T^2M=MT^2\ ,\eqno(2.5a)$$
where of course $T^2=\overline{T}^2$ here.
The other difference is that there is a vector $t\in\G^*_k$ in terms of which
the charge of the edge excitation $\la\in \pP^k$ is given by $t\cdot \la$.
For quantum Hall fluids we should have $M$ commuting
with the matrices $U_t$ and $V_t$ defined by
$$\eqalignno{(U_t)_{\la,\mu}=&\,\delta_{\la,\mu}\,\exp[2\pi \i t\cdot\la]&
(2.5b)\cr
(V_t)_{\la,\mu}=&\,\delta_{\la+t,\mu}\,\exp[-\pi \i{\rm Re}(\tau)\, t^2-2\pi \i 
{\rm Re}(t\cdot z)]\ .&(2.5c)\cr}$$
Physically, $U_t$ says that edge excitations should have integer total
charge, while $V_t$ is related to spectral flow. There are other properties
that $\overline{\z}$ is expected to obey in order for the theory to have a 
chance at being physical [4], but these are all which will be considered here.

Most observed plateaux lie in the Jain series with filling factor $\nu=
m/(ms\pm 1)$, for $s$ even. One intriguing explanation of those involves
the $W_{1+\infty}$ minimal models [3], but unfortunately these do not
possess [4] a modular invariant partition function in the sense given here
and so have an unclear RCFT interpretation. Instead, the partition functions
considered here correspond to `generic' $W_{1+\infty}$ RCFTs. The Jain
series can be obtained, both in the generic $W_{1+\infty}$ theories and
in the abelian Chern-Simons theories, when the $U_m$ algebra extends to 
$U_1\oplus\hat{A}_{m-1,1}$,
where $\hat{A}_{m-1,1}$ is affine $A_{m-1}$ at level 1. Thus we also
would like to know the modular behaviour of the characters of $\hat{
A}_{m-1,1}$. These turn out (Thm.\ 13.8 of [12]) to be given by the complex 
conjugates of the 
$\overline{S}$ and $\overline{T}^2$ matrices for $U_1$ at level $k=m$ (up to 
an irrelevant 
constant factor in the ${T}^2$ matrix). In particular the $\hat{A}_{m-1,1}$
weight $\la=\L_i$ corresponds to the $U_1$ weight $\la=i$.
Hence the weak invariants $M$ for  $U_{m,k}\oplus \hat{A}_{n,1}$ are in
natural one-to-one correspondence with the weak invariants $\overline{M}$
for $U_{m+1,k\oplus (n+1)}$, with the correspondence given by
$$M_{\la,\L_i;\mu,\L_j}=\overline{M}_{\la,{j};\mu,{i}}\eqno(2.6)$$
(note the $i\leftrightarrow j$ switch on the right side). 

The other algebra we are interested in is $\hat{A_1}$. The level $k$ here
is a nonnegative integer, and its level $k$ weights can be taken to be
the set $P^k_+=\{0,1,\ldots,k\}$. Its characters $\chi_a^k(\tau,z,u)$ can
also be expressed using theta functions (Ch.\ 13 of [12]), and its modular 
matrices are
$$\eqalignno{S_{a,b}=&\,\sqrt{{2\over k+2}}\sin[\pi{(a+1)\,(b+1)\over k+2}]\ ,
&(2.7a)\cr
T_{a,b}=&\,\delta_{a,b}\,\exp[\pi \i\{(a+1)^2/2(k+2)-1/4\}]\ .&(2.7b)\cr}$$
In particular, the set of highest weights for $\hat{A}_{1,k}\oplus U_{m,\ell}$
is $P_+^k\times \overline{P}^\ell$, and the modular matrices are
$S\otimes \overline{S}$, $T^2\otimes \overline{T}^2$. The relation between the
$N=2$ super-Virasoro algebra at $c={3k\over k+2}$, and $\hat{A}_{1,k}\oplus
U_{1,4}\oplus U_{1,2k+4}$, is given at the start of Section 6.

\bigskip\bigskip\centerline{{\bf 3. The classification of $U_m$ modular
invariants}}\bigskip

Throughout this paper we use the convenient
notation $(x;y):=(x,\sqrt{-1}y)$ for any vector lying in the pseudo-Euclidean
vector space
$\R\otimes(\G_k;\G_k)$, where likewise $(\Lambda_1;\Lambda_2)$ denotes the
indefinite lattice $\Lambda_1\oplus\sqrt{-1}\Lambda_2$.\medskip

{{\it Theorem 1.}} 

\item{{\it (a)}} The set of all weak invariants $\overline{\z}$ (defined
after (2.3)) is in a natural one-to-one
correspondence with all self-dual $2m$-dimensional lattices $\L$ containing
$(\G_k;\,\G_k)$; all these will automatically obey (2.5a). 
When each $k_{ii}$ is even, $\overline{T}$ exists and
the physical invariants $\overline{\z}$ (defined after (2.3)) correspond to 
these $\L$ which are in addition even.

\smallskip
\item{{\it (b)}}  Choose any $t\in \Gamma^*_k$ and any
weak invariant $\overline{
\z}$, and let $\L$ be the corresponding lattice. Then $\overline{\z}$ commutes
with $U_t$ iff it commutes with $V_t$, iff $(t;t)\in\L$.\medskip

In particular, the partition function $\overline{\z}$ of the theory is 
proportional to the indefinite theta function 
$$\theta_\L((z;z)|\tau)=\sum_{(x_L;x_R)\in \L}\exp[\pi \i\tau x_L^2-
\pi \i\tau^*x_R^2+2\pi \i(z\cdot x_L-z\cdot x_R)]\eqno(3.1a)$$
of the lattice $\L$, and the coefficient matrix $M$ in (2.2a) is given by
$$M_{\la,\mu}=\left\{\matrix{1&{\rm if}\ (\la;\mu)\in\L\cr 0&{\rm otherwise}
\cr}\right. \ .\eqno(3.1b)$$
In the following section we discuss how to find these lattices $\L$.

An alternate, level-independent, formulation of this classification using
Grassmannians is also possible and very intriguing. Let $Z_o^m$ be the
set of all weak invariants for $U_m$, for arbitrary level $k$, which 
are  not physical (i.e.\ violate (2.3b)), and let $Z_e^m$ be the physical
invariants for $U_m$. These
then correspond to odd (resp.\ even) self-dual lattices $\L$, by the
correspondence of Theorem 1(a). Up to transformations in the full orthogonal
group $SO(m,m)$, these lattices are unique: $I_{m,m}=(\Z^m;\Z^m)$
(resp.\ $II_{m,m}=II_{1,1}^m$ where $II_{1,1}$ has basis $\{(e/\sqrt{2};
\pm e'/\sqrt{2})\}$). Recall [13] that the Grassmannian $G_{m,n}(\R)=SO(m+n)/
(SO(m)\times SO(n))$ is an $mn$-dimensional  compact symmetric space consisting 
of all $m$-dimensional subspaces of $\R^{m+n}$. Its dual $G_{m,n}(\R)^*=
SO(m,n)/(SO(m)\times SO(n))$ is noncompact and consists of all $m$-dimensional
Euclidean subspaces in the pseudo-Euclidean space $\R^{m,n}$. By a rational 
point in $G_{m,n}(\R)^*$ we mean an equivalence class containing a rational 
matrix, or
equivalently a subspace $V$ with a basis $\{f_1,\ldots,f_m\}$ which can
be written over $\Q$ in terms of the preferred orthonormal basis $\{e_1,
\ldots,e_m,e_1',\ldots,e_n'\}$ of $\R^{m,n}$. Now the group
$SO(m,m)$ acting on our lattices $\L$ will  mix their left- and right-sectors
and hence change the physics. On the other hand, the group $SO(m)$ acting
separately on either side should preserve the physics. Hence we get a
natural bijection between the physically distinct $\overline{\z}\in Z_o^m$ or
$\overline{\z}\in Z_e^m$, and the set 
of rational points on $G_{m,m}(\R)^*$ (or between numerically distinct
$\overline{\z}$'s in $Z_o^m$ or $Z_e^m$, and rational points on $SO(m,m)$). 
In particular, the important 
lattices $\L_L$, $\L_R$ defined shortly are given by $\L_L=V\cap I_{m,m}$
and $\L_R=\sqrt{-1}(V^{\perp}\cap I_{m,m})$ for all odd $\L$, and $\L_L=V\cap 
II_{m,m}$ and $\L_R=\sqrt{-1}(V^{\perp}\cap II_{m,m})$ for even $\L$.

What makes this picture interesting is that it gives many examples 
of what the `moduli space' of certain families of RCFTs looks like.
In particular we find that (at least as far as their genus 1 partition
functions are concerned) the Wess-Zumino-Witten models corresponding
to $\widehat{u(1)^{\oplus_m}}$ form a dense subset of a noncompact
$m^2$-dimensional symmetric space.

It is intriguing that the norm condition (2.5a) is redundant here. 
A more striking example of the irrelevance of (2.3b) or (2.5a)
is provided in Thm.\ 2 below.

Part {\it 1(a)} of the theorem gives the classification of all partition 
functions for RCFTs corresponding to $U_m$ at any level $k$. The only other 
such classifications for all levels of an affine algebra are $\widehat{A_1}$,
$\widehat{A_1\oplus A_1}$, and $\widehat{A_2}$ (see [6] for the original
references). In Section 5 we generalise Thm.\ 1 to the algebra $\hat{A}_1
\oplus U_{m}$ at any level.

Part {\it 1(b)} of the theorem gives the complete classification of the effective
field theories for quantum Hall fluids, assuming they possess a partition
function $\overline{\z}$ discussed in the previous section. The reason for 
believing they should is given in [4]. As mentioned earlier,
this includes all generic (as opposed to minimal) $W_{1+\infty}$ theories, 
and all abelian  Chern-Simons theories considered in e.g.\ [2]. 

The relation between $\L$ (or $\overline{\z}$) and the physical quantities of the
quantum Hall fluid are  discussed in e.g.\ [2,4]. For example, the 
dimensionless Hall conductivity is
$\sigma_H=t\cdot t$. In all cases the relevant level is not $k$, corresponding
to the lattice $\G_k$, but rather the matrices $k_L$ and $k_R$ corresponding 
to the largest $m$-dimensional sublattice $\L_L:=\G_{k_L}$ of $\L$, and  
$\L_R:=\Gamma_{k_R}$ of $\sqrt{-1}\L$, which contains $\G_k$. That is, we are
interested in the `maximally extended chiral algebras' of the theory, rather
than the arbitrarily chosen subalgebra at level $k$. (In general it is a
very difficult problem to find the maximally extended chiral algebras for
an RCFT, but for $U_m$ theories it is trivial.) As an example, the Wen 
topological order
gives the degeneracy of the quantum Hall ground state on compact genus
$g$ surfaces, and will equal $|\L_L|^g$, as can be seen
directly from Verlinde's formula (this is discussed in [4]).

Because of this remark about chiral algebras, the $\overline{\z}$'s in Theorem 1
will include redundancies caused by an inappropriate original choice of level 
$k$ (incidently, these redundancies are avoided in the Grassmannian picture).
 To avoid these redundancies, it suffices to restrict attention to those
$\L$ with $\G_k=\L_L$. But, in order to keep all the $\z$'s obtained in
Theorem 1, we are then required to allow `heterotic' theories, i.e.\
theories whose `left level' $k_L$ need not equal its `right level'
$k_R$. In order to avoid the redundancies spoken of earlier, we would then
supplement the conditions of the previous section with one more:
$$M_{\la,0}=\delta_{\la,0}\ \quad{\rm and}\quad  M_{0,\mu}=\delta_{\mu,0}\ ,
\qquad\forall \la\in \pP^{k_L},\ \mu\in \pP^{k_{R}}\ .\eqno(3.2a)$$
In order for solutions $\overline{\z}$ to exist, it is necessary and 
sufficient to require that [14]
$$\L_L^*/\L_L\cong \L_R^*/\L_R\qquad{\rm and}\qquad \Q\otimes\L_L=
\Q\otimes\L_R\ .\eqno(3.2b)$$
The first condition is the isomorphism of groups, and is required by the
maximality property of $\L_L$ and $\L_R$ [14]. It says among other things
that $|\L_L|=|\L_R|$ (see (4.1a)). The second statement
states that $\L_L$ and $\L_R$ are rationally equivalent, and because
$\L_L$ and $\L_R$ are integral is equivalent to the existence of 
$2m$-dimensional self-dual lattices $\L$  containing $(\L_L;
\L_R)$. ([14] gives a practical algorithm for deciding when two lattices
are rationally equivalent.) These two conditions are independent: e.g.\ 
$\L_L=\Z\oplus\sqrt{3}\Z$
and $\L_R=A_2$ obey the first condition but not the second. It seems 
heterotic theories may not be directly physically relevant
here, because the corresponding partition function will not be real. We will 
not consider this redundancy issue again in this paper, and will not impose
(3.2a) (until Example 2 in the next section).

Heteroticity applies also to $U_t,V_t$: in general these will be replaced
by $U_{t_L}$, $U_{t_R}$, etc. The $U$-commutativity constraint then 
becomes
$U_{t_L}M=MU_{t_R}$, and Thm.\ 1(b) then becomes that $U$-commutativity
is equivalent to $(t_L;t_R)\in\L$ (heterotic $V$-commutativity is more
complicated to interpret because of its $z$-dependence, but would 
require at least that $t_L^2=t_R^2$ in order to be equivalent to 
$U$-commutativity). If we insist that $t_{L,R}$ satisfy
$$x_L\cdot t_L-x_R\cdot t_R\equiv x_L^2-x_R^2\qquad({\rm mod}\ 2)\qquad
\forall (x_L;x_R)\in\L\ ,\eqno(3.3a)$$
then we can say much about $t_{L,R}$. They always exist (it is easy to
see that if $\{e_1,\ldots,e_n\}$ is any basis of $\L$ and $\{e_1^*,\ldots,
e_n^*\}$ is the dual basis, then
$$(t_L;t_R)=\sum_{i=1}^ne_i^2\,e_i^*\eqno(3.3b)$$
satisfies (3.3a)). Although there is no unique solution $t_{L,R}$ to (3.3a)
(if $(t_L;t_R)$ works, so will anything in $2\L+(t_L;t_R)$),
the physically important quantities $t_{L,R}^2$ are severely constrained. For 
example, $t_{L,R}^2$ will be an integral multiple of $1/|k_{L,R}|$ 
(since $t_{L,R}$ must lie in $\L_{L,R}^*$). Moreover, any $t_{L,R}$
satisfying (3.3a) will obey
$$t_L^2\equiv t_R^2\qquad ({\rm mod}\ 8)\ .\eqno(3.3c)$$

The physical invariants for $\hat{A}_{m-1,1}$
are classified in [15]. [4] uses this analysis to find many (but not all) 
weak invariants for $U_2$ obeying (2.5) for the choice $t=e_1^*$.
[4] also found several (but not all) weak invariants for $U_2$ (ignoring
$U$-,$V$-commutativity). Some $\overline{\z}$ 
for the most physically interesting theories were found first in [5].

The theories corresponding to chiral quantum Hall lattices [2] are a
small subset of the theories in {\it 1(b)}. In particular, they correspond to
the special cases where $M$ in (3.1b) is the identity matrix. [4] explains
why the other $M$ also seem interesting and should be considered:
they could describe simple theories corresponding to some
experimentally observed plateaux not lying in the Jain sequences.
We are not claiming however that our theorem trivialises in any way the work 
in e.g.\ [2]. They are really  addressing the formidable task of finding 
explicit lists of those $\overline{\z}$ in {\it 1(b)} lying within the 
subclass of interest to them. As will be
described in the next section, this is so challenging that it is hopeless in 
general, but is possible if one restricts to sufficiently small $k$ and $m$, 
as they do.

The remainder of this section is devoted to a proof of the theorem. The
argument closely follows the one given in Lemma 3.1 of [6], and is 
surprisingly simple.

Note first that (2.3a),(2.4b) and the unitarity of $\overline{S}$ implies
$$M_{\la,\mu}=\sum_{\alpha,\beta\in \pP^k} \overline{S}_{\la,\alpha}\,
M_{\alpha,\beta}\,\overline{S}_{\beta,\mu}^{\ *}={1\over|k|}\sum_{\alpha,\beta\in 
\pP^k}\exp[2\pi \i(\mu\cdot \beta-\la\cdot\alpha)]\,M_{\alpha,\beta}\ .
\eqno(3.4a)$$
Taking absolute values and using the triangle inequality, (3.4a) becomes
$|M_{\la,\mu}|\le |M_{0,0}|$ with equality, thanks to (2.2b), iff 
the following  holds:
$$M_{\alpha,\beta}\ne 0\qquad \Longrightarrow\qquad \la\cdot \alpha\equiv
\beta\cdot\mu\quad({\rm mod}\ 1)\eqno(3.4b)$$
for all $\alpha,\beta\in\pP^k$. We know then that each $M_{\la,\mu}\in
\{0,1\}$. Define a set $\L$ by
$$\L=\bigcup_{{\la,\mu\in\pP^k \atop M_{\la,\mu}=1}}
(\la;\mu)+(\G_k;\G_k)\ .\eqno(3.4c)$$
Then (3.4b) implies that $\L$ is closed under addition and under multiplication
by $-1$, and therefore is a lattice. Also, (3.4b) says that whenever $x,y\in
\L$, then $x\cdot y\in\Z$ -- i.e.\ $\L$ is integral.
Putting $\la=\mu=0$ in (3.4a) says $\|\L/(\G_k
;\G_k)\|=|k|=|\G_k|$ and hence that $|\L|=1$ (see
(4.1a) below). Thus $\L$ is self-dual.

The rest of the theorem now follows quickly. 
Commutation with $U_t$ says that $(t;t)\in\L^*$, and while commutation
with $V_t$ says $(t;t)+\L=\L$. Hence both are equivalent to $(t;t)\in\L$.

\bigskip\bigskip\centerline{{\bf 4. Finding the self-dual lattices $\L$}}
\bigskip

In this section we address the question of finding all the self-dual $\L$
occurring in Theorem 1, i.e.\ making the classification of the partition
functions $\overline{\z}$ somewhat more explicit. It would seem however that
{\it this problem is completely intractible} for large $m$, simply because
the number of such $\L$ becomes so great. For example it includes, as
a small subset, the classification of all Euclidean self-dual lattices
of dimension $m$, and though there are only 28 of these for $m=20$, there
are over $8\times 10^{16}$ for $m=32$ (see e.g.\ Tables 2.2 and 16.3 of [11]).
 Also, we learned in the
last section that our lattices $\L$ for fixed $m$ (and varying level)
form a dense subset of an $m^2$-dimensional manifold!
These considerations give some indication of the numbers of $\overline{\z}$'s 
involved.
But apparently this is not a serious issue, because stability considerations
[2] seem to require small $m$ and $k$.

A point worth repeating is that, up to transformations in the
orthogonal group $SO(m,m)$, each $\L$ is equivalent either to the lattice 
$I_{m,m}$ (if odd) or the lattice $II_{m,m}$ (if even).
However $SO(m,m)$ mixes up
quite thoroughly the excitations on the two edges, and so those transformations
will not respect the physics in any way. 
On the other hand, transformations from the smaller group $SO(m)\times SO(m)$
should preserve the physics, and we will usually identify lattices related by
such transformations.

At least for small $m$, lattices are easy to work with and are conducive
to explicit computations. We begin this section with some general statements
[14] about how to find these $\L$, given $\G_k$, and then we specialise to $m
\le 2$. A basic
geometrical fact, easily provable by considering volumes of fundamental
regions, is the following: if $\L_1\subset \L_2$ are two integral lattices, 
then
$$\|\L_2/\L_1\|=\sqrt{|\L_1|/|\L_2|}\ ;\eqno(4.1a)$$
also, $\L_2^*/\L_2$ must be a subgroup of $\L_1^*/\L_1$, and $\L_2$ a
sublattice of $\L_1^*$.

The first step to solving our problem consists of finding all possible
$m$-dimensional integral lattices $\L_L$ which contain $\G_k$. 
Any $m$-dimensional integral
$\L_L$ containing $\G_k$ can be written as
$$\L_L=\G_k+\Z g_1+\cdots+\Z g_m \ ,\eqno(4.1b)$$
where $g_i\in \Gamma_k^*/\G_k$ obey $g_i\cdot g_j\in\Z$ (any or all $g_i$
may be 0).
Thus the task of finding all possible $\L_L$ reduces
to a finite search ($\|\G^*_k/\G_k\|=|k|$). 

Now choose any two such $\L_L$, and call the second one $\L_R$. We may or
may not have $\L_L=\L_R$, but we must have 
(3.2b). Let $h_1,\ldots,h_n$ be linearly independent generators of the group 
$\L_L^*/\L_L$. Find some $h_i'\in \L_R^*/\L_R$ such that $h_i
\cdot h_j\equiv h_i'\cdot h_j'$ (mod 1).
Again this is a finite search. Now define
$$\L=(\L_L;\L_R)+\Z(h_1;h_1')+\cdots+\Z(h_n;h_n')\ .\eqno(4.1c)$$
Such a $\L$ will be self-dual and contain $(\G_k;\G_k)$, and
all such $\L$ can be obtained in this way.

This manner of constructing lattices is called `gluing' (see e.g.\ Chapter
4 of [11]). There is another
standard method, called `shifting' [14], which is more elegant in some 
respects. We will only state a special case of it here.
Let $\L$ be a self-dual lattice, $V=\{v_1,\ldots,v_n\}\subset \Q\otimes \L$,
with each $v_i\cdot v_j\in\Z$. Define
$$\L(V):=\bigl\{x+\sum\ell_iv_i\,|\,\ell_i\in
\Z,\ x\in\L,\ {\rm and}\ \forall j,\ x\cdot v_j\in\Z\bigr\}.\eqno(4.2)$$
Then $\L(V)$ will also be self-dual.

\medskip\noindent{{\it Example 1.}}\quad $g=\widehat{u(1)}$ {\it at level 
$k\in\Z_{\ge}$} 

Here $\G_k=\sqrt{k}\,\Z$. The possible $\L_L$ are given by $d/\sqrt{k}\,\Z$ 
where $d\in\Z_{\ge}$ obeys
$$d|k\ ,\quad k|d^2\ .\eqno(4.3a)$$
Here $\L_L=\L_R$ is forced, by $|\L_L|=|\L_R|$. Now choose any
$\ell\in\{1,\ldots,{d^2/k}\}$ obeying 
$$\ell^2\equiv 1\qquad ({\rm mod}\ d^2/k)\ .\eqno(4.3b)$$
To any such $(\ell,d)$ there corresponds a distinct self-dual lattice 
$\L_{d,\ell}$ given by
$$\L_{d,\ell}:=(d/\sqrt{k}\,\Z\,;d/\sqrt{k}\,\Z)+\Z(\sqrt{k}/d\,;\ell\sqrt{k}
/d)\ , \eqno(4.3c)$$
and hence a weak invariant $\overline{\z}_{d,\ell}$. Conversely, any weak
invariant for $U_1$ at level $k$ is of this form.

A simple counting argument shows that there is exactly one such partition
function $\overline{\z}$ for each divisor of $k$ if $k$ is odd, or for each 
divisor
of $2^{a-2}k$ when $k$ is even, where $2^a$ is the exact power of 2 dividing
$k$. For example, for $k=1,2,\ldots,10$ there are precisely 1, 1, 2, 3, 2, 2, 2,
5, 3, 2 different $\L$'s, respectively. When $k$ is odd, we can 
make this
correspondence explicit using shifting (4.2): to any divisor $d$ of $k$, it 
is given by $d\mapsto  \L_1(\{(\sqrt{k}/d;-\sqrt{k}/d)\})$, where
$\L_1=(\sqrt{k}\Z;\sqrt{k}\Z)+\Z(1/\sqrt{k};1/\sqrt{k})$.

The relationship between the notation here and that of equation (4.26) of 
[4] is $k\mapsto p$, $\ell\mapsto \omega^{-1}$ (or $\omega_i^{-1}$
if $k$ is even), and $d\mapsto p/\delta$ (or $p/\delta'$ if $k$ is even).
However their list appears to
miss some $\overline{\z}$. For example, for $k=8$, they get six $\overline{\z}$'s, 
but two of
them are redundant. There are in fact five distinct solutions -- they miss 
the one with $d=4$. In general they will miss some $\overline{\z}$ when $k$ 
is even.

If we consider even $k$ and impose the stronger condition (2.3b), we find 
that the resulting lattices are in a one-to-one relationship with divisors of
$k/2$. This result was first obtained in [16].

If we impose commutation with $U_t$ for $t=1/\sqrt{k}$, then of course 
only one solution survives: $\ell=1$, $d=k$.

\medskip\noindent{{\it Example 2.}}\quad $g=\widehat{u(1)\oplus u(1)}$

It is difficult and unenlightening to state the $m=2$ solution for general 
$k$, although the list of $\L$ is easy to find for fixed $k$.
Instead we will give all the self-dual lattices $\L$ with 
$|k_L|\le 10$. For convenience we will mod out by $SO(2)\times SO(2)$.
Table 15.1 of [11] is a list of all 2-dimensional integral
lattices $\G_k$ of small determinant. These lattices give the possible values
of levels $k$, through the correspondence $\Gamma_k\mapsto k$. To avoid 
unnecessary redundancy, by level here we will mean
the minimal possible, namely $k_{L,R}$ -- i.e.\ we impose (3.2a).
A priori, the two levels $k_L,k_R$ need not be equal, but
for the small determinants considered here, (3.2b) usually forces them to be.
For convenience here, we will give components of weights in terms of the
$e_i$, not $e_i^*$ as before. Recall that $e_i\cdot e_j=k_{ij}$. We will
also give for each of these $\L$ the smallest values of $t_{L,R}^2$ for 
$t_{L,R}$ satisfying (3.3a).

\item{$|k|=1$.} The only choice of level is $k=\left(\matrix{1&0\cr0&1}
\right)$. The only $\L$  here is $I_{2,2}$.  $t_{L,R}^2=2$ is the
smallest.

\item{$|k|=2$.} Here $k=\left(\matrix{1&0\cr0&2}\right)$. The weights 
$\la\in\pP^k=\G_k^*/\G_k$ are generated linearly by $h=(0,{1\over 2})$. 
There is 
only one $\L$ here, corresponding to the `diagonal glue' $(h;h)$ (see (4.1c)).
$t_{L,R}^2=1$ is minimal.

\item{$|k|=3$.} There are two possibilities here: $k'=\left(
\matrix{1&0\cr0&3}\right)$ and $k''=\left(\matrix{2&1\cr1&2}\right)$. Each
of these produce exactly one $\L$, in both cases corresponding
again to the diagonal glue, for the generators $h'=(0,{1\over 3})$
and $h''=({1\over 3},-{2\over 3})$, respectively. Minimal $t'_{L,R}{}^2$,
$t''_{L,R}{}^2$ are ${4\over 3}$ and 0, respectively.

\item{$|k|=4$.} Again there are two possibilities: $k'=\left(
\matrix{1&0\cr0&4}\right)$ and $k''=\left(\matrix{2&0\cr0&2}\right)$.
The generators here are $h'=(0,{1\over 4})$, and $h''_1=({1\over 2},0)$
and $h''_2=(0,{1\over 2})$ respectively. There are two $\L$
for $k'$, but one of them involves a level reduction to $|k|=1$ and so
will be discarded. The other corresponds to the diagonal glue $(h';h')$.
There is only one $\L$ for $k''$, corresponding to the diagonal glues
$(h''_1;h''_1)$ and $(h''_2;h''_2)$. Minimal $t_{L,R}'{}^2,t_{L,R}''{}^2$
are 1 and 0.

\item{$|k|=5$.} We have $k'=\left(\matrix{1&0\cr0&5}\right)$ and
$k''=\left(\matrix{2&1\cr1&3}\right)$. $h'=(0,{1\over 5})$, and $h''=
({1\over 5},-{2\over 5})$. Again, there is only one $\L$
each, corresponding in each case to the diagonal glue, with minimal 
$t_{L,R}^2$ being ${6\over 5}$ and ${2\over 5}$, resp.

\item{$|k|=6$.} We have $k'=\left(\matrix{1&0\cr0&6}\right)$   and
$k''=\left(\matrix{2&0\cr0&3}\right)$, with $h'=(0,{1\over 6})$, and
$h''_1=({1\over 2},0)$ and $h''_2=(0,{1\over 3})$. There is a unique
$\L$ for each choice, again given by the diagonal glues, with minimal
$t_{L,R}^2$ being 1 and ${1\over 3}$.

\item{$|k|=7$.} We have $k'=\left(\matrix{1&0\cr 0&7}\right)$ and
$k''=\left(\matrix{2&1\cr 1&4}\right)$, where $h'=(0,{1\over 7})$ and
$h''=({1\over 7},-{2\over 7})$. The diagonal glues have minimal $t_{L,R}^2$ 
equal to ${8\over 7}$ and 0.
Here for the first time we have a heterotic possibility: $k_L=k'$, 
$k_R=k''$, with $(h';2h'')$ as the glue and minimal $t_{L,R}^2= {8\over 7}$.
The remaining lattice has $k_L=k''$, $k_R=k'$, with glue $(h'';3h')$ and
$t^2_{L,R}={8\over 7}$.

\item{$|k|=8$.} There are three possible levels here: $k'=\left(
\matrix{1&0\cr 0&8}\right)$, $k''=\left(\matrix{2&0\cr 0&4}\right)$, and
$k'''=\left(\matrix{3&1\cr 1&3}\right)$. The generators are: $h'=(0,
{1\over 8})$; $h''_1=({1\over 2},0)$ and $h''_2=(0,{1\over 4})$;
$h'''=({1\over 8},-{3\over 8})$. The choice $k'$ for the level yields
two $\L$'s, one with the diagonal glue $(h';h')$ (with $t_{L,R}^2=1$) and the
other with $(h';3h')$ (with $t_{L,R}^2={3\over 2}$). The choice $k''$ also has 
two, but one reduces to 
$|k|=2$ so can be ignored. The other is given by the diagonal glues (with 
$t_{L,R}^2=0$).
The final choice $k'''$ also has two $\L$'s, corresponding to $(h''';h''')$
($t_{L,R}^2={1\over 2}$) and $(h''';3h''')$ ($t_{L,R}^2={1\over 2}$).

\item{$|k|=9$.} There are three levels here as well: $k'=\left(\matrix{1&0
\cr0&9}\right)$; $k''=\left(\matrix{3&0\cr 0&3}\right)$; and 
$k'''=\left(\matrix{2&1\cr 1&5}\right)$. The generators are: $h'=(0,
{1\over 9})$; $h_1''=({1\over 3},0)$ and $h''_2=(0,{1\over 3})$; 
$h'''=({1\over 9},-{2\over 9})$. There is one $\L$ for each 
choice of level, and each is given by the diagonal glues (a second $\L$ 
for $k'$ reduces to $|k|=1$ and so is ignored) (with $t_{L,R}^2={10\over 9}$,
${2\over 3}$, and ${2\over 9}$ resp.).

\item{$|k|=10$.} We have $k'=\left(\matrix{1&0\cr 0&10}\right)$ and
$k''=\left(\matrix{2&0\cr 0&5}\right)$, with $h'=(0,{1\over 10})$,
$h''_1=({1\over 2},0)$, and $h''_2=(0,{1\over 5})$. As usual, there is
exactly one $\L$ for each level, and it corresponds to the
diagonal glues, and $t_{L,R}^2=1,{1\over 5}$, respectively.

In summary, there are exactly 1, 1, 2, 2(+1), 2, 2, 4, 5(+1), 3(+1), 2,
respectively, distinct self-dual lattices $\L$ (hence partition functions 
$\overline{\z}$)    for each $|k|\le 10$. Obviously this example
can be pushed considerably further.

An additional two-dimensional example is $U_1\oplus \hat{A}_{m-1,1}$, where
we impose $U_t$-commutativity
with $t=e_1^*$. As mentioned in [4], these will be given by tensor products
of the diagonal $\overline{\z}$ for $U_1$, with the various weak invariants 
for  $\hat{A}_{m-1,1}$. The latter are completely classified in Example 1.

\bigskip\bigskip\centerline{{\bf 5. Extensions}}\bigskip

A considerable amount of attention in the literature has been paid to the
classification of modular invariants for affine algebras ${g}$ -- see
e.g.\ [7,17,6] and references therein. Two of the more useful and general
(i.e.\ valid for any RCFT) concepts that have come from this  
are {\it simple  currents} and a certain {\it Galois action}.

A simple current -- see e.g.\ [17] -- can be defined as any label
$a\in P_+$ for which $S_{0,a}=S_{0,0}$ (compare (2.1c)).
 From Verlinde's formula one then finds 
that to any such $a$ corresponds a distinct permutation $J_a$ of $P_+$ and a  
function $Q_{J_a}:P_+\rightarrow \Q$, such that $J_a0=a$ and [17] 
$$S_{J_ab,c}=\exp[2\pi \i Q_{J_a}(c)]\,S_{b,c}\ .\eqno(5.1a)$$
From this fundamental equation can be derived (see Lemma 3.1 in [6], though 
the arguments are similar to that of Thm.\ 1 given above) the following 
important facts, valid for any weak invariant $M$ and any simple currents
$J,J'$:
$$\eqalignno{M_{J0,J'0}&\in\{0,1\}\ ;&(5.1b)\cr
M_{J0,J'0}=1 &\Longrightarrow \,M_{Ja,J'b}=M_{a,b}\qquad \forall a,b\in P_+\ 
;&(5.1c)
\cr M_{J0,J'0}=1&\Longleftrightarrow\ Q_J(a)\equiv Q_{J'}(b)\ ({\rm mod}\ 1)\ 
\forall\, a,b\in P_+\ {\rm with}\ M_{a,b}\ne 0\ .&(5.1d)\cr}$$
Let $\I$ denote the set (in fact, abelian group) of all simple currents
of $P_+$. An important subset of $P_+$ are the {\it fixed points} of $\I$,
defined by $a= Ja$ for some nonzero $J\in\I$.

The Galois action also concerns the matrix $S$. Verlinde's formula implies
[18] that each
entry $S_{a,b}$ will lie in some cyclotomic extension $K$ of $\Q$. Choosing
any Galois automorphism $\sigma\in{\rm Gal}(K/\Q)$, one finds [18]
$$\sigma S_{a,b}=\epsilon_\sigma(a)\,S_{a^\sigma,b}\eqno(5.2a)$$
for some map $\epsilon_\sigma:P_+\rightarrow \{\pm 1\}$ and permutation
$\la\mapsto \la^\sigma$ of $P_+$ (compare (5.1a)). This together with (2.3a) 
and (2.2b) implies
$$M_{a,b}=\epsilon_\sigma(a)\,\epsilon_\sigma(b)\,M_{a^\sigma,b^\sigma}\ .
\eqno(5.2b)$$
The most important consequence of (5.2b) is that, because of (2.2b), we get 
the selection rule
$$M_{a,b}\ne 0\quad\Longrightarrow\quad \epsilon_\sigma(a)=\epsilon_\sigma(b)
\quad \forall \sigma\in{\rm Gal}(K/\Q)\ .\eqno(5.2c)$$

\medskip\noindent{{\it Example 3.}}\quad ${g}=U_m$

Here every $\la\in \pP^k$ is a simple current, with $J_\la\mu=\la+\mu$ and
$Q_\la(\mu)=-\la\cdot\mu$. There are no fixed points. That $\pP^k$ consists 
only of simple currents is
precisely the reason the classification of $U_m$ weak invariants is so 
easy. For this reason their
classification would also follow from the work in [19]. By way of comparison,
[20] concerns the next simplest such class of algebras, $\widehat{A_1^{\oplus 
m}}$, which turns out to be far more complicated.

The cyclotomic field $K$ here can be taken to be $\Q(\zeta_{|k|})$, where
$\zeta_n:=\exp[2\pi \i/n]$, and all $\epsilon_\sigma(\la)=+1$. Gal$(\Q(
\zeta_n)/\Q)$ can be identified with the multiplicative group $\Z^*_n$
consisting of the integers $\ell$ mod $n$ coprime to $n$. We will write this
correspondence as $\sigma\leftrightarrow \sigma_\ell$. For $\ell$ coprime
to $|k|$, $\sigma_\ell\lambda=\ell\lambda$ is the Galois action for $U_m$.

\medskip\noindent{{\it Example 4.}} ${g}=\hat{A_1}$

Here there is exactly one simple current $J$, and it maps $a$ to $k-a$ and has 
$Q_J(a)=a/2$. The only fixed point is $k/2$. Write $\k=k+2$. The field $K$
 here can be taken to be $\Q(\zeta_{8\k})$. For any $\ell$ coprime to $8\k$,
$$\epsilon_\ell(a)=\left({2\k\over \ell}\right)\cdot\left\{\matrix{+1&{\rm 
if}\ \langle\ell{a+1\over 2\k} \rangle<{1\over 2}\cr -1&{\rm otherwise}\cr}
\right. ,\eqno(5.3a)$$
where $\langle x\rangle$ denotes the unique number between 0 and 1 congruent
to $x$ (mod 1). The factor $\left({2\k\over \ell}\right)$ is a Jacobi
symbol and, since it is independent of $a$, is irrelevant here. Also,
$$\sigma_\ell a=\left\{\matrix{2\k\,\langle{\ell\,(a+1)\over 2\k}\rangle&
{\rm if}\ \epsilon_\ell(a)=+1\cr 2\k-2\k\,\langle{\ell\,(a+1)\over 2\k}
\rangle&{\rm if}\ \epsilon_\ell(a)=-1\cr}\right. \ .\eqno(5.3b)$$

It is possible to completely solve the constraint (5.2c) for $\hat{A_1}$
at any level $k$. We find, provided gcd$(a+1,b+1,\k)=1$, that the right-side 
of (5.2c) is equivalent to $b\in\{a,Ja\}$, with the following exceptions:
$$\eqalignno{k=4:&\qquad a,b\in\{0,\,2,\,4\}\ ;&(5.4a)\cr
k=8:&\qquad a,b\in\{0,\,2,\,6,\,8\}\ ;&(5.4b)\cr
k=10:&\qquad a,b\in\{0,\,4,\,6,\,10\}\ ;&(5.4c)\cr
k=28:&\qquad a,b\in\{0,\,10,\,18,\,28\}\ \quad{\rm or}\quad
a,b\in\{6,\,12,\,16,\,22\}\ .&(5.4d)\cr}$$

When the gcd condition is not satisfied, simply divide through by the
common divisor (so $a'+1={a+1\over d}$, $b'+1={b+1\over d}$, $k'+2={\k\over d}$),
in order to apply this result. This result is actually far stronger than we
need -- in general it is necessary only to look at (5.2c) for $a=0$, which
for $\hat{A}_1$ was solved in Lemma 3 of [20], but our general solution (5.4) 
is an easy
consequence (sketch: we can assume $\k$ is even, $a$ odd, $b$ odd; if some 
prime $p$ divides gcd($a+1,\k$) but not $b+1$, then it can be shown that
$p$ must equal 3; thus gcd$(b+1,\k)=1$ and hence $b$ can get mapped by
(5.3b) to 0).

The physical invariants for $\hat{A_1}$ were first classified by [7] using
methods considerably different from the ones we use in this paper. The
newer techniques permit for example the following interesting generalisation 
of their important work:\medskip

{{\it Theorem 2.}} The list of all weak invariants
of $\hat{A_1}$ at level $k$ is:
$$ \eqalignno{{\cal A}_k=&\sum_{a=0}^k|\chi_a|^2\ ,\qquad\qquad\forall k; 
&(5.5a)\cr
{\cal D}_k=&\sum_{a=0}^k \chi_a\chi^*_{J^aa}\ ,\qquad\qquad\forall\, {\rm even}
\ k;&(5.5b)\cr
{\cal D}'_k=&\sum_{a=0}^{[(k-2)/4]}|\chi_{2a}+\chi_{k-2a}|^2+
\left\{\matrix{
2|\chi_{k/2}|^2&{\rm if}\ 4|k\cr 0&{\rm otherwise}}\right.\ ,\qquad
\forall\,{\rm even}\ k;&(5.5c)\cr
{\cal E}_4=&\,|\chi_0+\chi_4|^2+\chi_2(\chi_0+\chi_4)^*+(\chi_0+\chi_4)
\chi_2^*\ ,\qquad\qquad\quad k=4;&(5.5d)\cr
{\cal E}_8=&\,|\chi_0+\chi_2+\chi_6+\chi_8|^2\ ,\qquad\qquad\qquad\qquad \qquad
\qquad k=8;&(5.5e)\cr
{\cal E}_{10}=&\,|\chi_0+\chi_{6}|^2+|\chi_4+\chi_{10}|^2+|\chi_3+\chi_7|^2\ 
,\qquad\qquad\qquad k=10; &(5.5f)\cr
{\cal E}_{10}'=&\,|\chi_0+\chi_4+\chi_6+\chi_{10}|^2 ,\qquad\qquad\qquad 
\qquad \qquad\qquad k=10;&(5.5g)\cr
{\cal E}_{16}=&\,|\chi_0+\chi_{16}|^2+|\chi_4+\chi_{12}|^2+
|\chi_6+\chi_{10}|^2&\cr&\,+(\chi_2+\chi_{14})\chi_8^*+\chi_8(\chi_2+
\chi_{14})^*+|\chi_8|^2\ ,\quad \qquad\qquad k=16;&(5.5h)\cr
{\cal E}_{28}=&\,|\chi_0+\chi_{10}+\chi_{18}+\chi_{28}|^2+|\chi_{6}+\chi_{12}+\chi_{16}
+\chi_{22}|^2\ ,\qquad k=28.&(5.5i)\cr}$$
\medskip 

This theorem is proved in the Appendix.
The weak invariants in equations (5.5) differ from the physical invariant 
list of [7], in which (2.3b)
was also imposed, only in that there are a few extra
exceptionals (namely $\E_4$, $\E_8$ and $\E_{10}'$),
and that $\D_k$ is defined now for $4|k$ and $\D_k'$ for
$4|(k-2)$. It is surprising how irrelevant $T$-invariance is for the $\hat{A}_1$
classification. Of these, only ${\cal E}_4$ and ${\cal E}_8$ violate 
$T^2$-invariance. The names ${\cal A}_k$, ${\cal D}_k$, etc here are introduced
purely by analogy with the A-D-E classification in [7] -- it would be
very interesting however if some similar interpretation of this list can be 
found.

A similar result to Theorem 2 can be expected at least for $\hat{A_2}$ --
(5.2c) has also been solved for it. More generally,
it is easy to show that there are only finitely many weak invariants
at each level $k$, for each affine algebra $\hat{g}$.\medskip

{\it Theorem 3.}\quad
The complete list of weak invariants for $\hat{A_1}\oplus U_m$ which obey
(2.5a), at level 
$(k,\ell)$ for any positive integer $k$ and any positive definite integer
matrix $\ell$, is:\smallskip

\item{({\it sc}).} `simple current invariants': these are given in (A.8b), (A.9c),
or are of the form $\D'_k\,\overline{\z}$ (equivalently the tensor product
of the corresponding matrices) for any $\overline{\z}$ in Thm.\ 1;

\item{$k=4$.} the `$\E_7$-type exceptionals' given in
(A.14);

\item{$k=10$.}  the exceptionals given by the product
$\z={\cal E}_{10}'\,\overline{\z}$,
as well as the exceptionals given by the matrix product $M=M'\,(M_{10}\otimes 
\overline{I})$, where $M'$ is any simple current invariant in (sc), and 
$\overline{I}$ is the identity matrix for $\pP^\ell$;

\item{$k=16.$}  the exceptionals $\z={\cal E}_{16}\,
\overline{\z}$;

\item{$k=28.$}  the exceptionals $\z={\cal E}_{28}\,
\overline{\z}$.\medskip

Thm.\ 3 is also proved in the Appendix. The only `new'
invariants here (i.e.\ ones which cannot be generated by standard simple
current techniques [19] from those of $\hat{A}_1$) are the $k=4$ exceptionals, 
the
simplest of which occur for $A_{1,4}\oplus U_{1,6}$ and $A_{1,4}\oplus U_{1,9}$. 
The constraint (2.5a) is imposed here to shorten the proof; if instead we
drop (2.5a) then we get additional exceptionals only at $k=4,8,10$, and these 
can be easily found using the methods of the Appendix. Note that
few of our $\z$ factorise completely into a product of $m+1$ $\z_i$'s
-- in fact about half fail to  factorise into an $\hat{A_1}$ part and
a $U_m$ part. This is characteristic of modular invariant classifications
for semi-simple algebras and unfortunately means that the semi-simple 
classifications do not reduce to the simple ones. In general, there are
many more physical (or weak) invariants, including exceptionals,
for semi-simple algebras than would be expected from the lists for simple ones.

\bigskip\bigskip\centerline{{\bf 6. The $N=2$ superconformal minimal models}}
\bigskip

The reason we can read off the $N=2$ rational minimal model classification
from Thm.\ 3 is because [21] gives a
description of the $N=2$ super-Virasoro algebra at $c=3(1-{2\over k+2})$ for
$k\in\Z_{>}$, in  terms of the coset
$(SU(2)_k\times U(1)_4)/U(1)_{2k+4}$, and [22] explains how to reduce
physical invariant classifications for cosets to those for semi-simple
algebras ($\hat{A_1}\oplus U_2$ in our case).

The partition function of a superconformal field theory will not be built
directly from the super-Virasoro characters, since the super-Virasoro
algebra contains fields of half-integer conformal dimension. We are
required here to use certain projections to split the characters into two 
parts ($\tilde{\chi}_c^{a(b)}$
and $\tilde{\chi}_c^{a(b+2)}$ in the notation below), in other words to consider
the possible spin-structures. $SL_2(\Z)$ mixes these
spin-structures (apart from the periodic-periodic one, which contributes
an additive constant -- the Witten index for the Ramond sector -- to the 
partition function and will be ignored).

The partition functions for $N=2$ minimal models are of the form  [9]
$$\widetilde{\z}=\sum_{a,a'\in P_+^k}\sum_{c,c'=0}^{2k+3} 
\widetilde{M}_{a,c;a',c'}\,\tilde{\chi}_c^{a(b)}\,\tilde{\chi}_{c'}^{a'(b')*}
\ ,\eqno(6.1a)$$
where the $\widetilde{M}$'s are non-negative integers, $\widetilde{M}_{0,0;0,0}
=1$, the $\tilde{\chi}$'s are the `half-characters' alluded to above, and 
$b=0$ or 1 depending on whether $a+c$ is even or odd (similarly for $b'$). 
This must be invariant under the full $SL_2(\Z)$.
As with the quantum Hall effect (see the comments after (2.3)), we should
either regard $\tilde{\z}$ as a function of additional variables $z$ (other 
than just $\tau$), or equivalently, formally assume that all $\tilde{\chi}$'s 
are distinct.

Many $\tilde{\z}$ can be found in the literature (see e.g.\ [10,9]), but
the complete list appears here for the first time. The structure of $N=2$
minimal models have been studied in e.g.\ [23]. It is shown in [24] that
any rational model of the $N=2$ super-Virasoro algebra is unitary and
hence is one of the minimal models given below (this surprising result
is in sharp contrast to the $N=0$ and $N=1$ cases).

[22] tells us how to interpret this classification in terms of the
$A_{1,k}\oplus U_{1,4}\oplus U_{1,2k+4}$ one: in particular, 
$\tilde{\chi}_c^{a(b)}$
has identical modular behaviour as $\chi_a^k \ch_b^4 \ch_c^{2k+4\ *}
+\chi_{Ja}^k \ch_{b+2}^4 \ch_{c+k+2}^{2k+2\ *}$ and hence the classification
of $\widetilde{\z}$ in (6.1a) is identical to that of the physical
invariants
$$\z=\sum_{a,a'\in P_+^k}\sum_{b,b'=0}^3\sum_{c,c'=0}^{2k+3} M_{a,b,c;a',b',c'}
\,\chi_a^k\,\ch_b^4\,\ch_c^{2k+4}\,\chi_{a'}^{k*}\,\ch_{b'}^{4*}\,
\ch_{c'}^{2k+4*}\ ,\eqno(6.1b)$$
subject to the additional conditions that
$$M_{J0,2,0;0,0,k+2}=M_{0,0,k+2;J0,2,0}=1\ .\eqno(6.1c)$$
The precise relation between $M$ and $\widetilde{M}$ is given by [22]
$$\widetilde{M}_{a,c;a',c'}=M_{a,b,c';a',b',c}\eqno(6.1d)$$
(note the $c\leftrightarrow c'$ switch), where $b,b'\in\{0,1\}$ are as
defined after (6.1a). The classification of these $\z$ is an elementary
application of Thm.\ 3 -- all we need to do is impose (2.3b) and
(6.1c).

For convenience write $\k$ for $k+2$.

\medskip{\it Theorem 4.}\quad The complete list of distinct physical 
invariants $\widetilde{M}$ for $N=2$ minimal models at level $k$ are given 
by (6.1d), for each of the following choices of $M$: \smallskip

\item{$k$ {\it odd}}: There is only one kind of $M$ here: its only nonzero 
entries are
$$M^o_{a,b,c;J^{a+b+c}a,2a+bw+2c,av+bv+2zc+2\ell v}=1\qquad \forall 
\ell\in\Z\ ,\eqno(6.2a)$$
for any $a,b,c$ provided $vc/\k\in\Z$,
where $w\in\{1,3\}$ is arbitrary, and $z\in\{1,2,\ldots,{v^2\over \k}\}$ and 
the divisor $v$ of $\k$ are any solutions to
$$(4z^2- 1)\k/v^2\equiv v^2/\k\equiv 0\qquad ({\rm mod}\ 1)\ .\eqno(6.2b)$$

\item{$k/2$ {\it odd}}: There are three kinds of $M$'s here: we will call them
$M^{2,0}$, $M^{2,1}$ and $M^{2,2}$. The nonzero entries of $M^{2,0}$ are
$$M^{2,0}_{a,b,c;J^\ell a,ax+2dy+2e+2\ell,zc+ayv+2mv}=1\qquad \forall \ell,m\in\Z\ ,
\eqno(6.3a)$$
for all $a,b,c$ provided $d:=vc/\k\in\Z$ and $e:=(b-a)/2\in\Z$,
where $x\in\{1,3\}$, $y=\k\,(z^2-1)/2v^2$, $z\in\{1,2,\ldots,
2v^2/\k\}$, and $v$ is a divisor of $\k/2$, such that
$$v^2/\k\equiv \k\, (z^2-1)/2v^2\equiv 0\qquad({\rm mod}\ 1)
\ .\eqno(6.3b)$$
The nonzero entries of $M^{2,1}$ are
$$M^{2,1}_{a,b,c;J^\ell a,ax+2d\,(z+1)+2e+2m+2\ell,va+zd\k/v+2mv}=1\qquad
\forall \ell,m\in\Z\ ,\eqno(6.3c)$$
for all $a,b,c$ provided $d:=(c-av)v/\k\in\Z$ and $e:=(b-a)/2$, where
$x\in\{1,3\}$, $z\in\{1,2,\ldots,2v^2/\k\}$, and $v$ is a divisor of $\k/2$,
such that 
$$1/2+v^2/\k\equiv \k \,(z^2-1)/2v^2\equiv 0\qquad({\rm mod}\
1)\ .\eqno(6.3d)$$
The nonzero entries of $M^{2,2}$ are
$$M^{2,2}_{a,b,c;J^\ell a,ax+2e+wd+2\ell,av(x-w)/2+cz+\k\ell+2vm}=1\qquad
\forall\ell,m\in\Z\ ,\eqno(6.3e)$$
for all $a,b,c$ with $d:=cv/\k\in\Z$ and $e:=(b-a-d)/2\in\Z$, where $x,w\in
\{1,3\}$, $z\in\{1,\ldots,2v^2/\k\}$, $\k/v$ is odd, and
$$v^2/\k\equiv\k\,(z^2-1)/4v^2\equiv 0\qquad({\rm mod}\ 1)\ .\eqno(6.3f)$$

\item{$k/2$ {\it even}}: There are four kinds of $M$'s here: $M^{4,0}$,
$M^{0,1}$, $M^{04,2}$, and $M^{04,3}$.
$M^{4,0}$ exists only for $k\equiv 4$ (mod 8). Its nonzero entries are
$$M^{4,0}_{a,b,c;J^{ax+bx'+d}a,ay+by'+2d(z+1)+2\ell,av+bv+dz\k/v+2\ell v}=1 
\qquad \forall \ell\in\Z\ ,\eqno(6.4a)$$
for all $a,b,c$ provided $d:=(c-vb)v/\k\in\Z$, where $y,y'\in\{1,3\}$,
$x={1\over 4}+v^2/2\k$, $z\in\{1,2,\ldots,2v^2/\k\}$, $x'=(y-y')/2$, and $v$ 
is a divisor of $\k/2$, such that
$$2v^2/\k\equiv \k\,(z^2-1)/v^2\equiv 0\qquad
({\rm mod}\ 1)\ .\eqno(6.4b)$$
$M^{0,1}$ exists only when 8 divides $k$. Its nonzero entries are
$$M^{0,1}_{a,b,c;J^\ell a,bk/4+dy,bxv/2+2cz+2mv}=1+\delta_{a,k/2}\qquad
\forall \ell,m\in\Z\ ,\eqno(6.4c)$$
for all $a,b,c$ provided $a$ is even and $d:=vc/\k\in\Z$, where
$x,y\in\{1,3\}$, $z\in\{1,2,\ldots,v^2/\k\}$, and $v$ is a divisor of $\k$,
such that
$$v^2/\k\equiv 2\k\,(4z^2-1)/v^2\equiv 0\qquad({\rm mod}\ 1)\ .\eqno(6.4d)$$
$M^{04,2}$ exists whenever $k/2$ is even. Its nonzero entries are
$$M^{04,2}_{a,b,c;J^\ell a,bx+2d(z+1)+2m,bv+dz\k/v+2mv}=1+\delta_{a,k/2}
\qquad \forall \ell,m\in\Z\ ,\eqno(6.4e)$$
for all $a,b,c$ provided $a$ is even and $d:=(c-bv)v/\k\in\Z$, where
$x\in\{1,3\}$, $z\in\{1,2,\ldots,2v^2/\k\}$, and $v$ is a divisor of $\k/2$
such that
$$2v^2/\k\equiv \k\,(z^2-1)/v^2\equiv 0\qquad ({\rm mod}\ 1)\ .
\eqno(6.4f)$$
$M^{04,3}$ also exists whenever $k/2$ is even. Its nonzero entries are
$$M^{04,3}_{a,b,c;J^\ell a,ax+bx+2\ell,byv+d\k z/v+2\ell v}=1\qquad\forall 
\ell\ ,\eqno(6.4g)$$
for all $a,b,c$ for which $d:=(c-bv)v/\k\in\Z$, where $x,y\in\{1,3\}$, 
$z\in\{1,2,\ldots,4v^2/\k\}$ and $v$ is a divisor of $\k/2$ such that
$$2v^2/\k\equiv \k\,(z^2-1)/4v^2\equiv 0\qquad({\rm mod}\ 1)\ .
\eqno(6.4h)$$

\item{$k=10$.} In addition to the ones mentioned in (6.3), there are precisely
20 exceptionals, given by the matrix product $M'\,({\cal E}_{10}\otimes \overline{I})$,
where $M'$ here is any of the 20 matrices (4 $M^{2,0}$'s and 16 $M^{2,2}$'s) 
in (6.3) for $k=10$.

\item{$k=16$.} In addition to the ones mentioned in (6.4), there are
precisely 30 exceptionals, given by the tensor product ${\cal E}_{16}\otimes
\overline{M}$, where $\overline{M}$ here is the projection to the last two 
components of any of the 24 matrices $M^{0,1}$ or 6 matrices $M^{04,2}$ for 
$k=16$ (the matrices of
type $M^{0,1}$ and $M^{04,2}$ in (6.4) are always of the form ${\cal D}_k'
\otimes \overline{M}$).

\item{$k=28$.} In addition to the ones mentioned in (6.4), there are
precisely 24 exceptionals, given by the tensor product ${\cal E}_{28}\otimes
\overline{M}$, where $\overline{M}$ here comes from the 16 matrices $M^{4,0}$
or the 8 matrices $M^{04,2}$ for $k=28$.\medskip

As usual for the classifications considered in this paper, there is an
unavoidable problem with being explicit, at least for general $k$.
The number of level $k$ $N=2$ minimal models for $k<30$, is: 4, 10, 4, 20,
4, 14, 6, 28, 4, 40, 4, 20, 8, 18, 4, 72, 4, 20, 8, 20, 4, 28, 6, 28,
8, 20, 4, 64, 4.

The often-claimed A-D-E pattern of the $N=2$ minimal model classification
is rather obscure from the standpoint of our theorem, and is at best
`one-to-many' (e.g.\ the exceptional $E_7$ corresponds to 30 different
partition functions). An example of its apparent inappropriateness is that 
the so-called A and D partition functions for $k\equiv 2$
(mod 4)  correspond to the value 0 and 1, respectively, of the seemingly
insignificant parameter $x$ in (6.3)!

The remainder of this section is devoted to the proof of Theorem 4. Note the
definitions of $\I_{L,R}(M)$, $\p_{L,R}(M)$ given in (A.2b),(A.2c). Note also 
that (6.1c), (5.1d), and (2.3b) force
$$\eqalignno{M_{a,b,c;a',b',c'}\ne 0&\Longrightarrow a+b+c'\equiv a'+b'+c
\equiv 0\qquad ({\rm mod}\ 2)&(6.5a)\cr
M_{a,b,c;a',b',c'}\ne 0&\Longrightarrow {(a+1)^2\over 4\k}+{b^2\over 8}+
{c^2\over 4\k}\equiv {(a'+1)^2\over 4\k}+{b'{}^2\over 8}+{c'{}^2\over 4\k}
\quad({\rm mod}\ 1)&(6.5b)\cr}$$

\smallskip\noindent{{\it 6.1. The automorphism invariants}}

These are the physical invariants obeying (A.1a), i.e.\ $M_{0,0,0;a,b,c}=
\delta_{a,0}\delta_{b,0}\delta_{c,0}$. We know $\phi(0,0,1)=(J^x0,y,z)$
for some $x,y,z$, hence by linearity  (A.8b) we have $\phi(0,0,\k)=
(J^{x\k}0,\k y,\k z)= (J0,2,0)$,
where the last equality holds by (6.1c). For even $k$, this is impossible,
and hence for these $k$ there are no automorphism invariants.
The automorphism invariants for odd $k$ turn out to be a special case
($v=\k$) of the treatment in the next paragraph.

\smallskip\noindent{{\it 6.2. The ADE$_7$ invariants}}

These are the physical invariants obeying (A.2a).
Consider first $k$ odd. We want $M_{J^a0,b,c;0,0,0}=1$ for some $a,b,c$.
(6.5) says $a=b=0$ and $c$ is even, hence $v=c/2$ satisfies (6.2b). We
can choose $v$ so that it divides $\k$. We find $\I_L(M)=\I_R(M)$
is forced here by the constraint (A.2d). There are no fixed points here and
so $M$ is given by (A.2f) for some $\phi$. The way to
show $M$ satisfies (6.2a) is simply to look at the possibilities for
$\phi(1,0,0)$, $\phi(0,1,0)$, and $\phi(0,0,\k/v)$, 
and to solve the various constraints coming from (A.2g) and (6.5).

Next consider $k/2$ odd. Again, there are no fixed points, so (A.2f) applies.
Note that if $(1,e,e')\in\p_L(M)$ for some even $e,e'$,
then $(1,e,e')\mapsto (*,*,o)$ where $o$ is odd, in which case
(6.5b) cannot be satisfied. This, with (6.5b), forces either $(J,2,0)\in\I_L
(M)$ or $(J,2,\k)\in\I_L(M)$. Consider first the former possibility. Then
$(0,0,\k)\in\I_R(M)$ by (6.1c), hence we find
$$\I_L(M)=\I_R(M)=\langle(J,2,0),\,(0,2u,2v)\rangle\ ,$$
where $u\in\{0,1\}$ and $v$ divides $\k/2$. These two possible values of $u$ 
should be treated 
separately, and produce the invariants $M^{2,u}$ of (6.3). The second
possibility for $\I_L(M)$ either reduces to the former, or we have 
$$\I_L(M)=\I_R(M)=\langle(J,2,\k),\,(0,2u,2v)\rangle\ ,$$
where either $\k/v$ is odd (if $u=0$) or $\k/2v$ is odd (if $u=1$). We
then find $u=1$ violates (6.5b). $u=0$ here produces $M^{2,2}$.

Finally, consider $k$ a multiple of 4. Suppose first there is nothing in
$\I_L(M)$ of the form $(J,*,*)$, i.e.\ $\I_L(M)=\langle(0,2u,2v)\rangle$.
Then as before neither $\I_{L,R}(M)$ will have fixed points (see the
argument in the appendix after (A.9a)), so $\phi$ exists. Now $u=1$ because 
otherwise  $x_0:=(1,0,0)$ and $x_1:=(0,1,0)$ will have
$y_0\cdot y_1\equiv {1\over 4}$ (mod ${1\over 2}$) by (6.5), where $\phi(x_i)
=(J^{a_i}i,y_i)$, which
violates (A.8a). Now $\phi(x_0)=(J^x 1,y,w)$ where $y,w$ are odd. This 
means
$(0,2,2w)\in\I_R(M)$ (by (A.8b), $\phi(2,0,0)=(2,2,2w)$ must lie in $\I_R(M) 
(2,0,0)$) and hence (counting powers of 2) (A.2d)
forces $\I_L(M)=\I_R(M)$ and $w=v$. It is now straightforward to
verify $M$ is given by (6.4a).

Next, suppose $(J,0,0)\in\I_L(M)$. Then $(J,0,0)\in\I_R(M)$, because
of the argument in the Appendix after (A.9a), so 
$\I_L(M)=\I_R(M)=\langle(J,0,0),\,(0,2u,2v)\rangle$.
$M$ factors in this case into $\D_k'\otimes \overline{M}$ (or
$\E_{16}\otimes \overline{M}$ for $k=16$),
so it suffices to find $\overline{M}$. As usual, the analysis 
depends on
whether $u=0$ or 1: the former case yields (6.4c) and the latter case
yields (6.4e).

The remaining possibility is that neither $\I_{L,R}(M)$ contain $(J,0,0)$,
but both contain something of the form $(J,u_{L,R},v_{L,R})$. Then
$v_{L,R}\ne 0$ by (6.5). Once again we find by the usual arguments that
$u_{L,R},v_{L,R}$ can be chosen so that
$\I_L(M)=\I_R(M)=\langle(J,2u,2v)\rangle$. We must have $u=1$, since otherwise
$v$ would be even and hence ${\k\over 2}\,(J,0,2v)=(J,0,0)$ would have to
lie in $\I_L(M)$, contrary to hypothesis. We find $M$ is given by (6.4g).

\smallskip\noindent{{\it 6.3. Exceptional levels}}

The only exceptional levels obeying (2.3b) are $k=10$ and 28 ($\Delta_{L,R}^2
\equiv {4\over 3}$ (mod 2) for $k=4$). Thm.\ 3 ($k=10,k=28$)
allows us to read off
the answer from the $M$ of (6.3) and (6.4c), (6.4e).

\bigskip\bigskip\centerline{\bf 7. Conclusion}\bigskip

In this paper we accomplish two main modular invariant partition function 
classifications:
that of the possible effective field theories for the quantum Hall fluids;
and that of the $N=2$ superconformal minimal models. The answer to the
former is given in Section 3 in terms of self-dual lattices, where we
also provide a prettier but less practical formulation of the classification
in terms of rational points on Grassmannians. The answer to the second
problem is given in Section 6. This latter classification
 is often claimed to fall into an
A-D-E pattern, but from the complete list of partition functions obtained here
this claim looks rather artificial -- e.g.\ arbitrarily large numbers of 
partition functions are assigned to the same name $A_\rho$ (or $D_\rho$)
for large $\rho$. Certainly it is far less convincing a match than the A-D-E
of the ${A}_1$ classification [7].

The connection between these two problems lies in their symmetry algebras:
$u(1)^{\oplus_m}$ versus $(A_{1,k}\oplus u(1)_4)/u(1)_{2k+4}$. Solving the
first takes us a long way towards solving the second. In fact, in Section
5 we find the partition functions
 for the algebra $A_1\oplus u(1)^{\oplus_m}$ -- the choice
$m=2$ then yields the $N=2$ classification.

The moduli space picture of rational points on the dual Grassmannian
$G_{m,m}(\R)^*$ is a very intriguing one, reminiscent of the compactification
of heterotic strings on tori studied in [25]. It should be possible to find
a natural lattice interpretation for the $A_1\oplus u(1)^{\oplus_m}$
classification given here, and from this perhaps an analogous description
of its moduli space.

An interesting consequence of the work here is the list of all $A_1$
modular invariants, when invariance under $\tau\mapsto \tau+1$ is dropped.
This is given in Section 5. What is found is the list is surprisingly
little changed from the A-D-E list of  Cappelli-Itzykson-Zuber. One is
not always interested in invariance under the full $SL_2(\Z)$ (witness
the quantum Hall effect; other examples are statistical models with a 
fault line whose position can be shifted by gauge transformations;
see also e.g.\ [26]), and at
least in this case the answer is little different and is achieved
by similar methods.

This paper suggests many future projects. The $U(1)^m$  partition functions
(especially the nondiagonal ones with small $m$ and $k$ missing from
the lists in [2,4]) should be
studied in connection with the physics of the quantum Hall effect.
The new $N=2$ minimal models can be explored within the context of
Gepner models, elliptic genus, Landau-Ginzburg theories, etc. The 
modular invariant partition functions for the (0,2) minimal models [27]
can also be classified using the methods of Section 6.

\bigskip\bigskip I appreciate the hospitality of the MPIM, and have 
benefitted from several useful
communications with Andrea Cappelli, 
Wolfgang Eholzer, Christoph Schweigert and Mark Walton.
The possibility of classifying the $N=2$ minimal models was first suggested to
me by Jean-Bernard Zuber.

\bigskip\bigskip\centerline{\bf Appendix. Proofs for Section 5}\bigskip

We begin the Appendix with a sketch of the arguments which we will use
for obtaining the classifications given in Section 5. First consider
any weak invariant $M$ satisfying 
$$M_{a,0}=\delta_{a,0},\quad M_{0,b}=\delta_{b,0},\qquad\forall a\in P_+^L,\
b\in P_+^R\ .\eqno(A.1a)$$
Such $M$ are called {\it automorphism invariants}. It is possible to show
(e.g.\ this is a special case of Lemma 3.1(b)(iii) and Lemma 3.2(b) in 
[6])  that for any such $M$, there exists a bijection $\phi:P_+^L\rightarrow
P_+^R$ such that
$$M_{a,b}=\delta_{b,\phi a}\ .\eqno(A.1b)$$
For example, $\phi 0=0$ by (2.2c). Then (2.3a) reduces to
$$S_{a,b}=S_{\phi a,\phi b}\qquad \forall a,b\in P_+^L\ .
\eqno(A.1c)$$
To find all such $M$, i.e.\ all such $\phi$, we follow the technique first
developed in [28]. In particular, let $G$ be any subset of $P_+^L$ with
the property that for any $a,b\in P_+^L$,
$$S_{c,a}/S_{0,a}=S_{c,b}/S_{0,b}\qquad \forall c\in G\quad\Longrightarrow  
\quad a=b\ .\eqno(A.1d)$$
Any such set $G$ is called a {\it fusion-generator} for $P_+^L$ -- e.g.\ for 
$U_{m,k}$
we can take $G$ to be any set linearly generating $\overline{P}^k$,
while for $g=A_{r,k}^{(1)}$ we can take $G$ to be the set of fundamental
weights $\{\omega^1,\ldots,\omega^{(r+1)/2}\}$. $M$ is uniquely determined 
by how $\phi$       acts on $G$. See [28] for details.

The next step consists of weakening the constraint (A.1a) to
$$M_{a,0}\ne 0\quad\Rightarrow\quad a\in\I_L0,\quad{\rm and}\quad M_{0,b}\ne 0
\quad\Longrightarrow\quad b\in\I_R 0\ ,\eqno(A.2a)$$
where $\I_{L,R}$ are the groups of simple currents in $P_+^{L,R}$, 
respectively.
Any such $M$ is called an ADE$_7$-invariant [6], since these are precisely
the physical invariants of $\hat{A}_1$ satisfying
(A.2a). Useful definitions are
$$\eqalignno{\I_L(M):=&\,\{J\in\I_L\,|\,M_{J0,0}\ne 0\}\ ,&(A.2b)\cr
\p_L(M):=&\,\{a\in P_+^L\,|\,\exists b\in P_+^R\ {\rm such\
that}\ M_{a,b}\ne 0\}\ ,&(A.2c)\cr}$$
and define $\I_R(M)=\I_L(M^t)$, $\p_R(M)=\p_L(M^t)$. 
In the special case of an ADE$_7$-invariant, Lemma 3.1(b) of [6] says that
$I_{L,R}(M)$ are subgroups of $I_{L,R}$ obeying 
$$\eqalignno{\|\I_L(M)\|=&\,\|\I_R(M)\|\ ,&(A.2d)\cr
\p_{L,R}(M)=&\,\{a\in P^{L,R}_+\,|\,Q_J(a)\in\Z\ ,\ \forall J\in\I_{L,R}(M)\,\}\ .
&(A.2e)\cr}$$
The notion of fusion-generator $G(\I')$ for a group $\I'$ of simple currents
can be defined analogously to (A.1d), but some extra care is required
(see Def.\ 3.3 of [6]). For $g=A_{1,k}$, $G(\{0,J\})=\{2\}$. Ideally,
no $a\in G(\I_{L,R}(M))=:G_{L,R}$ will be a fixed point of
$\I_{L,R}(M)$, and for each $a\in G_L$  we will have $M_{a,f}=0$ for all
fixed points $f$ of $\I_R(M)$ (and similarly for each $b\in G_R$). When
this happens, the situation turns out to resemble the automorphism
invariant one: there will exist a bijection $\phi:\p_L(M)/\I_L(M)
\rightarrow \p_R(M)/\I_R(M)$ such that 
$$M_{a,b}={\|\I_L(M)\|\over \sqrt{\|\I_L(M)a\|\,\|\I_R(M)b\|}}\,
\delta_{\I_R(M) b,\phi(\I_L(M)a)}\ ,\eqno(A.2f)$$
and again this $\phi$ is uniquely determined by its values on $G_L$. 
 Moreover, if neither $a$ nor $b$ are fixed points of $G_L$,
$$S_{a,b}=S_{\phi a,\phi b}\ .\eqno(A.2g)$$
 (When possible, we will often write $\phi(a)$ for any element of $\phi(\I_L
 (M)a)$.) This is Lemma 3.3(b) of
[6]. For example, $\phi(\I_L(M))=\I_R(M)$. In order to prove that this `ideal
case' holds for a given choice of $P_+^{L,R}$, one needs constraints
on when $M_{a,f}\ne 0$ for $f$  $(a)$  (not) a fixed point of $\I_R(M)$
$(\I_L(M))$, respectively. In this paper we are only interested in the case 
$\|\I_L(M)\|=2$, in which case
$$M_{a,f}\ne 0\quad \Longrightarrow \quad S_{0,f}/S_{0,a}\in\{1,2\}\ .
\eqno(A.2h)$$
This is proved by evaluating (2.3a) at $(a,0)$ and $(0,f)$.

The final step in these classifications is to consider arbitrary weak
invariants $M$ and solve the constraints for those $b\in P_+^R$ satisfying
$M_{0,b}\ne 0$. One constraint is given by (5.2c) with $a=0$. Another
useful constraint is [20]
$$\eqalignno{\sum_{a\in P_+^R}M_{0,a}\,S_{a,b}\ge&\, 0\qquad \forall b\in
P_+^R\ ,&(A.3a)\cr
\sum_{a\in P_+^R}M_{0,a}\,S_{a,b}=&\, 0\qquad \Longleftrightarrow
\qquad b\not\in \p_R(M)\ .&(A.3b)\cr}$$
Of course similar equations hold for $P_+^L$ and $\p_L(M)$. These are
proved by evaluating (2.3a) at $(0,a)$ and using (2.1c) and (2.2b).
These are severe constraints and we find that for almost all $M$, (A.2a)
will be satisfied.\medskip

Now let us turn to the proof of Thm.\ 2. Let $M$ be any weak invariant
for $A_{1,k}$. Recall the discussions about $A_{1,k}$ at the end of Section 
2 and in Example 2. Write $\k:=k+2$. The automorphism invariants are
easy to find: (A.1c) and (2.7a) say
$$\sin(2\pi/\k)=\sin(\pi\,(\phi 1+1)/\k)\ ,$$
and hence $\phi 1\in\{1,J1\}$. $S_{1,1}=S_{\phi 1,\phi 1}$ says
$\phi 1=J1$ is only possible when $k$ is even. Since $G=\{1\}$, we are now
done: we find $M={\cal A}_k$ if $\phi 1=1$, and $M=\D_k$ if $\phi 1=J1$.

Next, consider the ADE$_7$-invariants which are not automorphism invariants.
Then (A.2d) says that $\I_L(M)=\I_R(M)=\{0,\,J\}$. 
By (5.1d), $k$ must be even. We choose the fusion-generator 
$G_{L,R}=\{2\}$. The only fixed point $f$
is $f=k/2$ (for odd $k/2$, $f\not\in\p_{L,R}(M)$ and can be ignored). We want 
to show
that $M_{2,f}=0$ (except for the trivial case where $k=4$, when
$2=f$). The only solution to (A.2h) is $k=16$, and $M_{2,f}\ne 0$ or $M_{f,2}
\ne 0$ for
$k=16$ is easily seen to yield $M=\E_{16}$ (see e.g.\ Section 7.2 of [6]).
Otherwise, $M$ will be `ideal'; in this case
$\phi(2)=2$ is forced by (A.2g) at (2,0), so uniqueness forces $M=\D'_k$.

Finally, consider an arbitrary weak invariant $M$ for $A_1$.
We learned in (5.4) that (A.2a) is forced, except possibly when
$k=4,8,10,28$. These succumb to a case-by-case analysis. 

Consider first $M$ for $k=4$ violating (A.2a). Then
by (5.4a), we may assume without loss of generality that
$M_{0,2}\ne 0$. Put $b=2$ in (A.3a) and use (5.4a) to obtain
$$\sin(3\pi/6)-M_{0,2}\sin(3\pi/6)+M_{0,4}\sin(3\pi/6)>0\ .\eqno(A.4)$$
Therefore $M_{0,2}\le M_{0,4}$ and hence $M_{0,2}=M_{0,4}=1=M_{4,0}=M_{4,4}$ 
by (5.1b), (5.2b) (see (5.3b)).
Computing (2.3a) at (0,0) now forces $M_{2,0}=1$. From (5.1d) we know that 
$M_{a,b}=0$ if either $a$ or $b$ 
is odd. That $M_{2,2}=0$ follows from (2.3b) at (0,2). Hence $M=\E_4$,
given in (5.5d).

The argument for $k=8$ is similar. Suppose $M_{0,2}+M_{0,6}\ge 1$. Using
(5.4b) and putting $b=4$ into (A.3a) gives
$$(1+M_{0,8}-M_{0,2}-M_{0,6})\,\sin(5\pi/10)\ge 0\eqno(A.5a)$$
while (2.3b) at (0,1) gives
$$(M_{1,1}+M_{7,1})\,\sin(2\pi/10)=(1-M_{0,8})\,\sin(2\pi/10)+(M_{0,2}-
M_{0,6})\,\sin(4\pi/10)\ .\eqno(A.5b)$$
(A.5b) forces $M_{0,2}=M_{0,6}$, since $\sin(4\pi/10)/\sin(2\pi/10)$ is
irrational. Then (A.5a) forces $M_{0,8}=1$. (5.2b) with $\sigma=\sigma_3$
then gives $M_{2,0}=M_{2,2}=1$. By (5.1d) and (A.3b), $\p_L(M)=
\p_R(M)=\{0,2,6,8\}$, so we are done by (5.1c).

For $k=10$, (5.4c) and $b=1,2,3$ in (A.3a) tells us
$$\eqalignno{1-M_{0,10}\ge&\,|M_{0,4}-M_{0,6}|&(A.6a)\cr
1+M_{0,10}\ge&\,M_{0,4}+M_{0,6}\ .&(A.6b)\cr}$$
The only difficult task here is eliminating the possibility $M_{0,10}=0$,
$M_{0,4}\ne 0$. In this case, $M_{0,4}=1$ and $M_{0,6}=0$. (5.2b) then implies 
$M_{a,0}=M_{0,a}$ for all $a$, so by (A.3b) we have $2\not\in\p_L(M)$. (2.3a) 
at (0,1) and (2,1) give $2M_{5,1}+M_{1,1}+
M_{9,1}=2$ and $M_{1,1}+M_{9,1}=M_{5,1}$ using (5.4c), i.e.\ $M_{5,1}=2/3$, 
which is impossible.

For $k=28$ use (5.4d) and $b=1,2,3$ in (A.3a), and then
 $\sigma=\sigma_7,\sigma_{11}$ in
(5.2b). The rest of the argument is as before. This concludes the proof of
Thm.\ 2.
\medskip

Now we turn to the classification of weak invariants $M$ for $U_{m,\ell}
\oplus A_{1,k}$ (Thm.\ 3). Though much more complicated notationally than for 
$\hat{A}_1$, and involving many more cases, the arguments are very similar to 
those used in Thm.\ 2. For later convenience we will replace the level
$\ell$ of $U_m$ with $\ell_{L,R}$, where $|\ell_L|=|\ell_R|$. 
The sets of  highest weights here are
$P_{L,R}=\pP^{\ell_{L,R}}\times P_+^k$.  The possible simple currents are $\I_{L,R}=
\pP^{\ell_{L,R}}\times\{0,J\}$. Let $\p_{L,R}'(M)$
 denote the projections of $\p_{L,R}(M)$ onto $\pP^{\ell_{L,R}}$. 

Any weak invariant $M'$ for $U_{m,\ell}$ and  $M''$ for $A_{1,k}$, gives us
a weak invariant $M=M'\otimes M''$ of $U_{m,\ell}\oplus A_{1,k}$. The
converse unfortunately is not true. We begin with the following useful
fact, true for any $g$ (not just $g=A_{1,k}$), which tells us when $M$
actually does factorise.

\medskip{\it Claim.}\quad Let $M$ be a weak invariant for $U_{m,\ell}
\oplus g$. Suppose that for each $x\in\p_L'(M)$, there exists an $x'\in 
\pP^{\ell_R}$ for which $M_{x,0;x',0}\ne 0$, and conversely that
for each $y\in\p_R'(M)$ there is a $y''\in \pP^{\ell_L}$ such that
$M_{y'',0;y,0}\ne 0$. Then $M=M'\otimes M''$ for some weak invariants
$M'$ and $M''$ of $U_{m,\ell}$ and $g$, resp.

\medskip{\it Proof.}\quad Define $M'_{x,y}:=M_{x,0;y,0}$, $M''_{a,b}
:=M_{0,a;0,b}$. We want to show 
$$M_{x,a;y,b}=M'_{x,y}\,M''_{a,b}\ .\eqno(A.7)$$
Suppose $M_{x,a;y,b}=0$. Then either $M'_{x,y}=0$, or $M_{x,0;y,0}\ne 0$
and (by (5.1c)) $M_{0,a;0,b}=0$ -- in either case (A.7) holds.

If instead $M_{x,a;y,b}\ne 0$, then again by (5.1c) and our hypothesis,
$M_{0,a;y-x',b}\ne 0$. Now consider any $M_{a,c;v,d}\ne 0$; by hypothesis,
there exists a $v''$ such that $M_{v'',0;v,0}\ne 0$, and hence by
(5.1d) $v\cdot (y-x')\in\Z$. Thus again by (5.1d) applied to $(0,0;y-x',0)$
and $(a,c;v,d)$, we must have $M_{0,0;y-x',0}=1$, i.e.\ $M_{x,0;
y,0}=1$. Then (5.1c) again forces (A.7).\qquad {\it
QED} \medskip

As before, consider first the  automorphism invariants $M$ obeying (A.1a).
(A.1c) with $a=(0,0)$ forces $\phi(x,c)\in\I_R(0,c)$. For a
fusion-generator choose $G=\{(x_1,0),\,\ldots,\,(x_n,0),$
$(0,1)\}$, where the $x_i$ span $\pP^{\ell_L}$. For each $i$, 
write $(y_i,J^{a_i}0): =\phi(x_i,0)$,
and also $(y_0,J^{a_0}1):=\phi(0,1)$. Then by (A.1c) these must obey
$$0\equiv x_i\cdot x_j-y_i\cdot y_j+ka_ia_j/2
\equiv y_i\cdot y_0-(a_i+ka_ia_0)/2\equiv y_0^2-ka_0/2\quad ({\rm mod}\ 1)
\ ,\eqno(A.8a)$$
for all $i,j\in\{1,\ldots,n\}$. By (5.1c) and the usual fusion arguments [28], 
we find
$$\phi(\sum_{i=1}^n c_ix_i,\,c_0)=(\sum_{i=0}^nc_iy_i,\,
\prod_{i=0}^nJ^{c_ia_i}\,c_0)\ .\eqno(A.8b)$$
It is straightforward to verify from (A.8a) that (A.8b) satisfies
(A.1c), and that $\phi$ is well-defined and  one-to-one. Hence it defines an 
automorphism
invariant, and all automorphism invariants are of this form.\smallskip

Next we consider the more general condition (A.2a). If both
$\I_{L,R}(M)\subset \pP^{\ell_{L,R}}\times\{0\}$, then this just reduces to
the automorphism invariant case considered in the previous paragraph:
`renormalise' the levels $\ell_{L,R}$ by replacing the lattices $\G_{L,R}$ 
with the denser lattices $\Gamma_{L,R}+\I_{L,R}(M)$. 

So assume $\I_R(M)\not\subset \pP^{\ell_R}\times\{0\}$. Again, by 
renormalising
the levels, we can require $\I_L(M)=\{0,\,(x_L,J^{a_L})\}$ and $\I_R(M)=
\{0,\,(x_R,J)\}$, where $2x_{L,R}=0$, $(x_L,J^{a_L})\ne 0$, and
$$x_L^2+ka_L/2\equiv x_R^2+k/2\equiv 0\quad ({\rm mod}\ 1)\ .\eqno(A.9a)$$
Suppose for contradiction that $x_R=0$ (so $k$ is even) but $x_L\ne 0$. Then 
we have a problem with (5.4), since
$1\in\p_L'(M)$ and $o\not\in\p_R'(M)$ for any odd
$o$. Therefore $x_L=0$ iff $x_R=0$.

So assume next that both $x_{L,R}\ne 0$. The fusion-generator $G_L$ for the 
left side can be chosen to be of the form $\{(x_1,0),\,\ldots,\,(x_n,0),\,
(x_0,1)\}$, where the $(x_i,0)$ generate all $(x,0)\in\p_L(M)$, and
$(x_0,1)\in\p_L(M)$.  For each $i\ge 0$,
choose some $y_i\in\pP^{\ell_R}$  such that
$$y_i\cdot x_R+\delta_{i,0}/2\equiv x_i\cdot x_j-y_i\cdot y_j\equiv 
0\quad ({\rm mod}\ 1)\ ,\eqno(A.9b)$$
for all $i,j\ge 0$. Define $M$ by (A.2f), with 
$$\phi(\sum_{i=0}^nc_ix_i,\,c_0)=(\sum_{i=0}^nc_iy_i,\,c_0)\ .\eqno(A.9c)$$
 Our $M$ must be of
this form, and the reader can readily verify that any such $M$ is a well-defined
weak invariant.

The final possibility for a weak  ADE$_7$-invariant here is that 
(after renormalising the levels) $\I_L(M)=\I_R(M)=\{0,\,(0,J)\}$. 
Assume first that $M_{x,0;y,k/2}=0$
for all $x,y$. Then for each $(x,0)\in\p_L(M)$ there exists a $x'$
such that $M_{x,0;x',0}\ne 0$. Using (5.1c), $x\mapsto x'$ must be a
bijection here ($|\ell_L|=|\ell_R|$), and so by the Claim $M$ factorises.

If instead $M_{x,0;y,k/2}\ne 0$ for some $x$, then (A.2h) 
forces $k=4$. For a given $x\in\pP^{\ell_L}$, it is easy to show (by 
evaluating (2.3a) at $(x,0;0,0)$ -- see also Lemma 3.2(b) of [6]) that there 
must exist some $y\in\pP^{\ell_R}$ such that either
$$\eqalignno{M_{x,0;z,a}=&\,\delta_{z,y}\,(\delta_{a,0}+\delta_{a,4})
&(A.10a)\cr
M_{x,0;z,a}=&\,\delta_{z,y}\,\delta_{a,2}\ .&(A.10b)\cr}$$
Let $\I_L(a)$ be the set of $x$ satisfying (A.10a), and $\I_L(b)$ be those
satisfying (A.10b) -- we know $\I_L(b)$ is nonempty
(otherwise $M$ would factorise).

Similarly, given any $y\in\pP^{\ell_R}$, there exists $x,x'\in\pP^{\ell_L}$ 
such that either
$$\eqalignno{M_{z,a;y,2}=&\,(\delta_{z,x}+\delta_{z,x'})\,(\delta_{a,0}+
\delta_{a,4})&(A.11a)\cr
M_{z,a;y,2}=&\,\delta_{z,x}\,(\delta_{a,0}+\delta_{a,4})+\delta_{z,x'}
\delta_{a,2}&(A.11b)\cr
M_{z,a;y,2}=&\,(\delta_{z,x}+\delta_{z,x'})\,\delta_{a,2}\ .&(A.11c)\cr}$$
Similarly put each such $y$ in $\I_R(A)$, $\I_R(B)$, $\I_R(C)$, respectively.
Of course similar remarks hold for $M^t$, giving us sets $\I_R(a),\ldots,
\I_L(C)$. 

Choose any $x\in\I_L(a)$, and $x_1,x_1',y_1$ satisfying any
of (A.11). Let $\Delta_L:=x_1-x_1'$. Then $\Delta_L\cdot x\in\Z$, by (5.1d).
Moreover, for any $x\in\I_L(b)$, (2.3a) evaluated at $(x,0;y_1,2)$
implies $x\cdot \Delta_L\in\pm {1\over 3}+\Z$. Since $\I_L(a)\cup\I_L(b)
=\pP^{\ell_L}$, what we have shown is that 
$$\eqalignno{\I_L(a)=&\,\{x\in\pP^{\ell_L}\,|\,x\cdot \Delta_L\in \Z\}&(A.12a)\cr
\I_L(b)=&\,\pm a_L+\I_L(a)\qquad{\rm for\ some}\ a_L\ .&(A.12b)\cr}$$
Of course there is a $\Delta_R$ playing the identical role for $\p_R(M)$.
Moreover, by (5.1d) and (A.12) we have
$$M_{x,a;z,c}\ne 0\ {\rm and}\ M_{y,b;z+j\Delta_R,d}\ne 0\quad
\Longrightarrow\quad x-y\in\{0,\pm\Delta_L\}\ ,\eqno(A.12c)$$
for any $x,y,z,a,b,c,d,j$ (similarly for $M^t$). From (5.1c) we see that
each nonempty $\I_{L,R}(A)$, $\I_{L,R}(B)$, $\I_{L,R}(C)$ can be written
as a union of cosets of $\I_{L,R}(a)$ in $\I_{L,R}$.

Let $x,x',y$ satisfy one of (A.11a) or (A.11c). Then $x'=x\pm \Delta_L$ for
some choice of sign, and hence by (2.5a)
$$\pm\Delta_L\cdot x\equiv\Delta_L^2\qquad({\rm mod}\ 1)\ .\eqno(A.13a)$$
If $x,x',y$ satisfy (A.11b), then similarly (2.5a) and (A.12c) say $x'=x\pm
\Delta_L$, and 
$$\pm\Delta_L\cdot x\equiv \Delta_L^2-1/3\qquad({\rm mod}\ 1)\ .\eqno(A.13b)$$ 

Suppose first that $\Delta_L^2\in\Z$. Then by (A.12a), $M_{\Delta_L,0;x_R,0}=
1$ for some $x_R$ -- by (A.12c) we can choose the sign of $\Delta_R$ so that
$x_R=\Delta_R$, hence also $\Delta_R^2\in\Z$. By (A.13), we must have 
$\I_{L,R}(A)=\emptyset$, $\I_{L,R}(C)=\I_{L,R}(a)$, $\I_{L,R}(B)
=\I_{L,R}(b)$. Let $x_1,\ldots,x_{n-1}$ be generators for $\I_L(a)$. Then for each
$i$ there is a $y_i$ such that $M_{x_i,0;y_i,0}=1$. Since $a_L\in\I_L(B)$,
there is a $b\in\I_R(B)$ such that $M_{a_L,2;b,2}=1$. Note that $x_1,\ldots,
x_n:=a_L,y_1,\ldots,y_n:=b$ satisfy (A.8a) with all $a_i=0$ and $y_0=0$. Hence
they define an automorphism invariant $M'$. Write $M''=MM'{}^{-1}$. It is easily
shown that its only nonzero entries are
$$M''_{x,J^i0;x,J^j0}=M''_{y,2;y,2}=M''_{y,J^i0;y-3(\Delta_L\cdot y)\Delta_L,
2}=
M''_{y,2;y+3(\Delta_L\cdot y)\Delta_L,J^i0}=M''_{x,2;x\pm\Delta_L,2}=1
\eqno(A.14a)$$
for all $x\in\I_L(a)$, $y\in\I_L(b)$, $i,j\in\{0,1\}$, and choices of signs
($M''_{x,2;x,2}=0$ follows from (2.3a) at $(a_L,0;x,2)$).
Conversely, any choice of $\Delta_L\in \pP^{\ell_L}$ with $3\Delta_L=0$ and
$\Delta_L^2\in\Z$, and any choice of an automorphism invariant $M'$ from
$\pP^{\ell_L}$ and $\pP^{\ell_R}$, defines a distinct weak invariant in this 
way, which obeys (2.5a).

Otherwise, $\Delta_L^2\not\in\Z$. By (A.13a) and (5.1c), $\I_{L,R}(C)=
\I_{L,R}(b)$, and hence (choosing signs appropriately) 
 $M_{0,2;\Delta_R,0}=M_{\Delta_L,0;0,2}=1$, so $\Delta_L^2\equiv
\Delta_R^2\equiv {2\over 3}$ (mod 1). By (A.13b), $\I_{L,R}(B)\subseteq
\I_{L,R}(b)$ and so $\I_{L,R}(B)=\emptyset$, $\I_{L,R}(A)=\I_{L,R}(a)$.
Again we get an automorphism invariant
$M'$ from (A.8), such that $M'':=MM'{}^{-1}$ has the following nonzero
entries:
$$M''_{x,J^i0;x,J^j0}=M''_{x\pm\Delta_L,J^i0;x,2}=
M''_{x,2;x\pm\Delta_L,J^j0}=M''_{x\pm\Delta_L,2;x\pm\Delta_L,2}
=1\ ,\eqno(A.14b)$$
for all $x\in\I_L(a)$.
Conversely, for any choice of $\Delta_L\in \pP^{\ell_L}$ with $3\Delta_L=0$
and $\Delta_L^2\equiv{2\over 3}$ (mod 1), and choice of automorphism invariant
$M'$ from $\pP^{\ell_L}$ to $\pP^{\ell_R}$, we get a distinct weak invariant
of this form which obeys (2.5a).

That exhausts all the weak ADE$_7$-invariants. Again the remaining exceptionals
will occur only at $k=4,8,10,28$. We will work out the case $k=10$ in
detail -- the remaining exceptional levels are easier, and succumb to similar 
arguments. Assume, by renormalising levels if necessary, that either 
$\I_L(M)=\I_R(M)=\{(0,0)\}$,
or $\|\I_{L,R}(M)\|=2$ and $(x,J)\in\I_R(M)$ for some $x$. 
Define $s_a:=\sum_xM_{0,0;x,a}$. Then $s_{10}\in\{0,1\}$ by (5.1b),
(5.1c). Putting $b=(0,1)$, (0,2), (0,3) in (A.3a) tells us
$$\eqalignno{1-s_{10}\ge &\,|s_4-s_6| &(A.15a)\cr
1+s_{10}\ge &\,s_4+s_6&(A.15b)\cr}$$
(compare (A.6)). Hence also $s_4,s_6\in\{0,1\}$. The argument against the 
possibility that $s_{10}=0$ but 
$s_4=1$ is similar to the analogous argument for $A_{1,10}$ (see below
(A.6)): again $(0,2)\not\in\p_L(M)$, so evaluating (2.3a) at (0,0;0,1) and (
0,2;0,1) gives $\sum_xM_{x,5;0,1}={2\over 3}$.

Next suppose $s_{10}=0$ but $s_6=1$. Then $\exists z_6,z_6'$ such that
$M_{0,0;z_6,6}=M_{z_6',6;0,0}=1$. Looking at each $b=(x,0)$ in (A.3a) forces 
$z_6'=0=z_6$. Thus $(x,a)\not\in\p_{L,R}(M)$ $\forall x$, for each 
$a=1,2,4,5,10$. We find using (5.4) that $M_{0,3;x,3}=
M_{0,3;x,7}=\delta_{x,z_3}$ for some $z_3$. We also find using (5.2b) that
$M_{0,4;x,4}=
M_{0,4;x,10}=M_{0,10;x,4}=M_{0,10;x,10}=\delta_{x,0}$. Let $x_i$ generate
$\pP^{\ell_L}$. Then for each $i$ there is a $y_i$ and $a_i$ such that 
$M_{x_i,0;y_i,J^{a_i}0}=1$. The conditions on $y_i$, $a_i$, and $y_0:=z_3$
coming from (5.1d) and (2.5a) are precisely  the congruences of (A.8a).
Hence there is an automorphism invariant $M'$ given by (A.8b), for which
$M=M'\,(\overline{I}\otimes \E_{10})$.

The remaining possibility is 
$$M_{0,0;0,0}=M_{0,0;z_4,4}=M_{0,0;z_6,6}=M_{0,0;z_{10},10}=1\eqno(A.16)$$
and all other $M_{0,0;x,a}=0$. Of course $z_4=z_{6}+z_{10}$ and $2z_{10}=0$.
$M^t$ will obey a similar equation, for parameters $z'_{4},z_6',z'_{10}$
(look at (2.3a) at (0,0;0,0)).
If $z_{10}=0$ then looking at (2.3a) at $(0,0;x,0)$ we see that also
$z_{10}'=0$ and the Claim applies and $M$
factorises. For $z_{10},z_{10}'\ne 0$, $\exists z_0$ such that $z_0\cdot z_{10}\equiv
{1\over 2}$ (mod 1). Looking at
(A.3a) with $b=(x,0)$ shows that either $z_6=0$, or $z_6=z_{10}$. But 
$z_6=z_{10}$
is ruled out by using an argument similar to that after (A.6b): we would have
$\sum_xM_{x,5;z_0,1}={4\over 3}$.

Thus $z_4=z_{10}\ne 0$, $z_6=0$. This succumbs to a similar argument to
the $s_{10}=0$, $s_6\ne 0$ one: we find the parameters (like $x_R:=z_{10}$)
obey constraints identical to (A.9a),(A.9b) (provided we impose (2.5a)), and 
hence defines a weak invariant
$M''$ such that $M=M''\,(\overline{I}\otimes\E_{10})$.

\bigskip\bigskip\centerline{{\bf References}}\bigskip

\item{[1]} R.\ E.\ Prange and S.\ M.\ Girvin, The Quantum Hall Effect,
2nd edition (Springer-Verlag, New York, 1990)

\item{[2]} J.\ Fr\"ohlich, U.\ M.\ Studer, and E.\ Thiran, A classification
of quantum Hall fluids, cond-mat/9503113 

\item{[3]} A.\ Cappelli, C.\ A.\ Trugenberger, and G.\ R.\ Zemba, 
Nucl.\ Phys.\ (Proc.\ Suppl.) B 45A (1996) 112 

\item{[4]} A.\ Cappelli and G.\ R.\ Zemba, Modular invariant partition
functions in the quantum Hall effect, hep-th/9605127

\item{[5]} M.\ Milovanovi\'c and N.\ Read, Edge excitations of paired
fractional quantum Hall states, cond-mat/9602113

\item{[6]} T.\ Gannon, Kac-Peterson, Perron-Frobenius, and the classification
of conformal field theories, q-alg/9510026

\item{[7]} A.\ Cappelli, C.\ Itzykson, and J.-B.\ Zuber, 
Commun.\ Math.\ Phys.\ 113 (1987) 1

\item{[8]} A.\ Cappelli, Phys.\ Lett.\ B185 (1987) 82

\item{[9]} D.\ Gepner, Nucl.\ Phys.\ B296 (1988) 757

\item{[10]} F.\ Ravanini and S.-K.\ Yang, Phys.\ Lett.\ B195 (1987) 202;

\item{} Z.\ Qui, Phys.\ Lett.\ B198 (1987) 497;

\item{} S.-K.\ Yang, Nucl.\ Phys.\ B285 (1987) 183;

\item{} {\it see also} J.\ Fuchs, A.\ Klemm, M.\ G.\ Schmidt and D.\
Verstegen, Int.\ J.\ Mod.\ Phys.\ A7 (1992) 2245

\item{[11]} J.\ H.\ Conway and N.\ J.\ A.\ Sloane,  Sphere Packings,
Lattices and Groups, 2nd edition (Springer-Verlag, Berlin, 1993)

\item{[12]} V.\ G.\ Kac, Infinite-dimensional Lie algebras, 3rd
edition (Cambridge University Press, Cambridge, 1990)

\item{[13]} R.\ V.\ Gamkrelidze (Ed.), Geometry I, Encyclopedia of Mathematical
Sciences Vol.\ 28 (Springer-Verlag, Berlin, 1991)

\item{[14]} T.\ Gannon and C.\ S.\ Lam, Rev.\ Math.\ Phys.\ 3 (1991) 331 

\item{[15]} C.\ Itzykson, Nucl.\ Phys.\ (Proc.\ Suppl.) 5B (1988) 150;

\item{} P.\ Degiovanni, Commun.\ Math.\ Phys.\ 127 (1990) 71

\item{[16]} D.\ Gepner and Z.\ Qui, Nucl.\ Phys.\ B285 (1987) 423

\item{[17]} A.\ N.\ Schellekens and S.\ Yankielowicz, Nucl.\ Phys.\ B327 
(1993) 673

\item{[18]} A.\ Coste and T.\ Gannon, Phys.\ Lett.\ B323 (1994) 316

\item{[19]} M.\ Kreuzer and A.\ N.\ Schellekens, Nucl.\ Phys.\
B411 (1994) 97

\item{[20]} T.\ Gannon, J.\ Math.\ Phys.\  36 (1995) 675

\item{[21]} P.\ Di Vecchia, J.\ L.\ Peterson, M.\ Yu and H.\ B.\ Zheng,
Phys.\ Lett.\ B174 (1986) 280

\item{[22]} T.\ Gannon and M.\ A.\ Walton, Commun.\ Math.\ Phys.\ 173 (1995)
175

\item{[23]} E.\ Kiritsis, Phys.\ Rev.\ D36 (1987) 3048;

\item{} G.\ Mussardo, G.\ Sotkov and M.\ Stanishtov, Phys.\ Lett.\ B218
(1989) 191

\item{[24]} W.\ Eholzer and M.\ R.\ Gaberdiel, Unitarity of rational $N=2$
superconformal theories, hep-th/9601163

\item{[25]} K.\ S.\ Narain, M.\ H.\ Sarmadi and E.\ Witten, Nucl.\ Phys.\ B279
(1987) 369

\item{[26]} I.\ G.\ Koh and P.\ Sorba, Phys.\ Lett.\ B215 (1988) 723

\item{[27]} P.\ Berglund, C.\ V.\ Johnson, S.\ Kachru and P.\ Zaugg,
Heterotic coset models and (0,2) string vacua, hep-th/9509170

\item{[28]} T.\ Gannon, Invent.\ Math.\ 122 (1995) 341

\end